\documentclass[twocolumn]{aastex631}
\usepackage{bm}
\usepackage{amsmath}
\usepackage{tabularx}

\newcommand{\vdust}{\bm{v}_\mathrm{d}}
\newcommand{\vgas}{\bm{v}_\mathrm{g}}
\newcommand{\vdx}{v_{\mathrm{d}x}}
\newcommand{\vdy}{v_{\mathrm{d}y}}

\newcommand{\vgx}{v_{\mathrm{g}x}}
\newcommand{\vgy}{v_{\mathrm{g}y}}

\newcommand{\alphaM}{\alpha_\mathrm{M}}
\newcommand{\rhog}{\rho_\mathrm{g}}
\newcommand{\rhod}{\rho_\mathrm{d}}
\newcommand{\st}{\operatorname{St}}
\newcommand{\Hgas}{H_\mathrm{g}}
\newcommand{\hgas}{h_\mathrm{g}}
\newcommand{\OmK}{\Omega_\mathrm{K}}
\newcommand{\taus}{\tau_\mathrm{s}}
\newcommand{\alphass}{\alpha_\mathrm{SS}}
\newcommand{\etaT}{\eta_{\mathrm{T}}}

\defcitealias{Hsu22}{HL22}
\defcitealias{Lin22}{LH22}

\begin{document}
\title{Streaming instabilities in accreting protoplanetary disks: A parameter study}

\author[0009-0005-0548-8791]{Shiang-Chih Wang}
\affiliation{Institute of Astronomy and Astrophysics, Academia Sinica, Taipei 10617, Taiwan}
\affiliation{Department of Physics, National Tsing Hua University, Hsinchu 30013, Taiwan}
\author[0000-0002-8597-4386]{Min-Kai Lin}
\affiliation{Institute of Astronomy and Astrophysics, Academia Sinica, Taipei 10617, Taiwan}
\affiliation{Physics Division, National Center for Theoretical Sciences, Taipei 10617, Taiwan}
\correspondingauthor{Shiang-Chih Wang, Min-Kai Lin}
\email{scwang@asiaa.sinica.edu.tw}
\email{mklin@asiaa.sinica.edu.tw}

\begin{abstract}
The streaming instability (SI) is currently the leading candidate for triggering planetesimal formation in protoplanetary disks. Recently, a novel variation, the `azimuthal-drift' streaming instability (AdSI), was discovered in disks exhibiting laminar gas accretion. Unlike the classical SI, the AdSI does not require pressure gradients and can concentrate dust even at low abundances. We extend previous simulations of the AdSI to explore the impact of dust abundance, accretion flow strength, pressure gradients, and grain size. For a dimensionless accretion flow strength $\alphaM=0.1$ and particle Stokes number $\st=0.1$, we find the AdSI produces dust filaments for initial dust-to-gas ratios as low as $\epsilon=0.01$. For $\epsilon\gtrsim 1$, maximum dust-to-gas ratios of order 100 are attained, which can be expected to undergo gravitational collapse. Furthermore, even in systems dominated by the classical SI, an accretion flow drives filament formation, without which the disk remains in a state of small-scale turbulence. Our results suggest that an underlying accretion flow facilitates dust concentration and may thus promote planetesimal formation. 
\end{abstract}

\section{Introduction} \label{sec:intro}

The leading hypothesis for the formation of km-sized or larger planetesimals --- the building blocks of planets --- is through the gravitational collapse of high-density dust or pebble clumps formed by the `streaming instability' \citep[SI,][]{Youdin05,Youdin07a,Johansen07} in protoplanetary disks (PPDs).

The SI stems from the mutual drag force between dust grains and their surrounding gas. Typically, PPDs exhibit an adverse pressure gradient, slightly reducing the gas rotation from pure Keplerian. Dust grains, on the other hand, rotate at the Keplerian speed and thus experience a headwind. Frictional drag removes angular momentum from the dust, causing it to drift inwards relative to the gas \citep{Whipple72,Weidenschilling77}. This relative drift provides the source of free energy for the SI. However, the precise instability mechanism has only recently been understood as a resonance between this relative drift and inertial waves in the gas \citep{Squire18,Squire20}.

Since its discovery, the SI has been studied extensively through direct numerical simulations \citep{Bai10a,Bai10b,Carrera15,Yang17,Flock21,Li21,Schafer22,Rucska23,Baronett24}. Modern, state-of-the-art 3D simulations with self-gravity show that the SI can indeed trigger planetesimal formation \citep{Simon16,Schafer17,Abod19,Lim23}, provided that the grain size or their abundance are sufficiently large \citep{Li21}. 

Most studies model the SI under idealized conditions: the disk is laminar, unmagnetized, isothermal, etc. It is necessary to relax these assumptions to assess the efficiency of the SI in a realistic PPD. There have been several efforts in this direction to account for, but not limited to: turbulence \citep{Gole20,Umurhan20}, magnetic fields \citep{Wu24}, non-isothermal gas \citep{Lehmann23}, multi-species dust \citep{Krapp19,Paardekooper20,Zhu21,Yang21}. In many of these cases, the SI is tempered. 

Another consideration is that PPDs are fundamentally accretion disks: gas drains onto the central star. It is unclear how the SI operates in tandem with such an accretion flow, which should be a basic feature of any realistic disk model. To this end, \cite[][hereafter LH22]{Lin22} analyzed the stability of dust-gas interaction in accreting PPDs. They found a new type of `azimuthal drift' streaming instability (AdSI) powered by the relative azimuthal drift between dust and accreting gas. They were motivated by the paradigm shift to magnetically-driven accretion (\citealp{Bai17,Bethune17,Suriano19,Hu22,Wang23,Hsu24}, see also reviews by \citealp{Lesur23,Pascucci23}). However, their model is agnostic to the exact accretion mechanism as they prescribe an external torque to drive it. Thus, the AdSI should apply to laminar accretion of other origins. 

A distinction between AdSI and classical SI is that the former can develop under vanishing radial pressure gradients. Indeed, in a direct follow-up, \cite[][hereafter HL22]{Hsu22} simulated the AdSI in models without a radial pressure gradient and found that dense dust filaments still form. Furthermore, the AdSI produces noticeable dust clumping even when the initial solid-to-gas mass ratio is less than unity. The AdSI may, therefore, be relevant to dust ring or planetesimal formation in disk regions with small metallicities and weak radial pressure gradients, where the classical SI is ineffective. 


However, \citetalias{Hsu22} only conducted selected simulations across a limited range in parameter space. This paper expands their work by simulating disks with various metallicities, accretion flow strengths, pressure gradients, and grain sizes. We quantify the conditions under which the AdSI can lead to strong clumping, examine how the AdSI behaves under dust poor conditions, and make direct comparisons with the classical SI. 


This paper is organized as follows. In \S\ref{sec:equation}, we describe the basic equations and parameters for modeling AdSI. We present our simulation method, setups, and diagnostics in \S\ref{sec:method}. Our results are presented in \S\ref{sec:result}, which include varying the initial dust-to-gas ratio and accretion flow (\S\ref{sec:vary_eps_alpha}); as well as varying grain size (\S\ref{sec:vary_St}), and varying initial radial pressure gradient (\S\ref{sec:varyeta}). We discuss our result in \S\ref{sec:disscussion} and summarize in \S\ref{sec:summary}.

\section{Disk Model and Parameters} \label{sec:equation}

We consider a PPD of gas and dust in orbit around a central star of mass $M_*$. We use cylindrical coordinates $(R,\phi,z)$ to denote the cylindrical radius, azimuth, and height centered on the star. The Keplerian frequency around the star is $\OmK=\sqrt{GM_*/R^3}$, where $G$ is the gravitational constant.


We use $(\rho_\mathrm{g}, P_\mathrm{g}, \bm{V}_\mathrm{g})$ to denote the density, pressure, and velocity of the gas. We assume a strictly isothermal gas with $P_\mathrm{g}=C_\mathrm{s}^2\rhog$, where $C_\mathrm{s}=\OmK H_\mathrm{g}$ is the constant sound-speed and $\Hgas$ is the pressure scale-height. The aspect ratio of the disk is then $\hgas\equiv H_\mathrm{g}/R$. The gas is also threaded by a magnetic field $\bm{B}$. However, its sole purpose in our model is to drive a background gas accretion and does not actively partake in the dynamics. This is expected to be valid under strongly non-ideal conditions as expected in PPDs. Following \citet[][see also \cite{McNally17}]{Hsu22}, we thus will account for the magnetic field as a prescribed body force in the hydrodynamic equations, described below. 

We approximate the dust grains as a single, pressureless fluid with density $\rhod$ and velocity $\bm{V}_\mathrm{d}$. We consider small grains with Stokes number $\st=\tau_\mathrm{s}\OmK \ll 1$, where the stopping time $\taus$ characterizes the frictional force between dust and gas, for which the fluid approximation of dust is applicable \citep{Jacquet11}. 

For simplicity, we neglect gas viscosity, dust diffusion, and self-gravity. 

\subsection{Shearing box approximation} \label{sec:basic_equation}
In this work, we study local dynamics with characteristic lengthscales $\ll R$. We thus model a small disk region by adopting the shearing box framework \citep{Goldreich65,latter17}. The center of the shearing box is placed at a point ($R_0, \phi_0, 0$) rotating around the star with angular frequency $\Omega_0=\OmK(R_0)$, and $\phi_0=\Omega_0 t$, where $t$ is the time. Cartesian coordinates ($x,y,z$) in the shearing box correspond to ($R,\phi,z$) directions in the global disk. In the local model, we can ignore global curvature and approximate Keplerian rotation as the linear shear flow $-(3/2)x\Omega_0\boldsymbol{\hat{y}}$. We then define $\bm{v}_\mathrm{d,g}=\bm{V}_\mathrm{d, g} + (3x/2 -R_0)\Omega_0\boldsymbol{\hat{y}}$ as the dust and gas velocities in the rotating frame relative to Keplerian shear. Hereafter, we drop the subscript $0$ for brevity. 

We also focus on dynamics near the disk midplane. We thus neglect the vertical component of the stellar gravity but retain the vertical dimension. This yields a vertically uniform disk in equilibrium. We assume axisymmetry throughout (thus $\partial_y \equiv 0$).

The axisymmetric, unstratified shearing box equations of dust and gas can be written as: 
\begin{align}
    &\label{eqn:gascontinuous}\frac{\partial\rhog}{\partial t} + \nabla \cdot(\rhog \vgas)=0,\\
    & \label{eqn:gasmomentum}
    \begin{aligned}
         \frac{\partial \vgas}{\partial t} + \vgas\cdot\nabla \vgas &= 2\vgy\Omega \boldsymbol{\hat{x}} - \vgx\frac{\Omega}{2}\boldsymbol{\hat{y}} - \frac{\nabla p}{\rhog}\\
         &+2\etaT R\Omega^2\boldsymbol{\hat{x}}+F_{\phi}\boldsymbol{\hat{y}} + \frac{\epsilon}{\taus}(\vdust-\vgas),
    \end{aligned}
    \\
    &\label{eqn:dustcontinuous}\frac{\partial\rhod}{\partial t} + \nabla \cdot(\rhod \vdust)=0,\\
    &\label{eqn:dustmomentum}\frac{\partial \vdust}{\partial t} + \vdust\cdot\nabla \vdust = 2\vdy\Omega \boldsymbol{\hat{x}} - \vdx\frac{\Omega}{2}\boldsymbol{\hat{y}} - \frac{\vdust-\vgas}{\taus}
\end{align}
\citepalias{Hsu22}, where $\epsilon=\rhod/\rho_\mathrm{g}$ is the local dust-to-gas density ratio. Here, $p$ is the local pressure fluctuation. The terms $\propto \taus^{-1}$ represent frictional coupling between dust and gas. 

In this work, $\epsilon$ represents the dust-to-gas ratio near the midplane of a stratified disk, towards which solids settle. This is related to the vertically-integrated dust-to-gas ratio, or metallicity $Z$, via $Z=\epsilon H_\mathrm{d}/H_\mathrm{g}$, where $H_\mathrm{d,g}$ are the dust and gas scale-heights, respectively. Based on our fiducial simulation, we estimate $H_\mathrm{d}\sim 0.01 H_\mathrm{g}$ \citep{Lin21}, which for $\epsilon=1$ gives $Z=0.01$, i.e., solar metallicity.


In Eq. \ref{eqn:gasmomentum}, we introduce a dimensionless total radial pressure gradient $\etaT$ (including gas and magnetic) and an azimuthal body force $F_{\phi}$ to represent the effect of global pressure gradients and magnetic fields, respectively. For a spiral field under ohmic resistivity, explicit expressions of $\etaT$ and $F_{\phi}$ can be found in \citetalias{Lin22}.

We define the reduced radial pressure gradient
\begin{equation}
    \widetilde{\eta}=\frac{\etaT}{\hgas}.
\end{equation}
The following simulations mostly consider $\widetilde{\eta}=0$, but we also briefly vary $\widetilde{\eta}$ to explore the behavior of SI. 

We parameterize $F_\phi$ through 
\begin{equation}
    \alphaM=-\frac{2RF_{\phi}}{C_s^2}.
\end{equation}
Here, $\alphaM$ characterizes laminar horizontal Maxwell stresses, which differ from turbulent mass and momentum diffusion measured in the simulations (see \S\ref{sec:parameter_study}). As shown below, the azimuthal body force, or torque, drives an equilibrium gas accretion. Our fiducial $\alphaM=0.1$.

The Stokes number $\st$ controls the degree of dust-gas coupling. To connect it to physical grain sizes, $a_p$ , we consider dust with internal density $\rho_{\bullet}=1 $ $\mathrm{g\,cm^{-3}}$ in the Minimum Mass Solar Nebula \citep[MMSN,][]{Hayashi81,Chiang10} under the Epstein drag regime to find 
\begin{equation}
    a_{\mathrm{p}}\simeq 100\left(\frac{\st}{0.1}\right)\left( \frac{R}{\mathrm{AU}} \right)^{-\frac{3}{2}}\ \mathrm{cm},
    \label{eqn:sttoparticle}
\end{equation}
\citep{Chen20}. Thus, $\st=0.1$, our fiducial value, corresponds to cm-sized grains in the outer disk at $\sim 10\mathrm{AU}$. Our fiducial initial dust-to-gas ratio is $\epsilon=1$. 

\subsection{Equilibrium state}

Eqs. \ref{eqn:gascontinuous}-\ref{eqn:dustmomentum} admit steady states with constant $\rho_\mathrm{g}$ and $\rhod$, with drift velocities:
\begin{align}
    \label{eq:vgx}
    &\frac{\vgx}{C_s}=\frac{1}{\Delta^2}\left[2\epsilon\st\widetilde{\eta}-(\st^2+\epsilon+1)\alphaM \hgas\right],\\ 
    \label{eq:vgy}
    &\frac{\vgy}{C_s}=-\frac{1}{\Delta^2}\left[(\st^2+\epsilon+1)\widetilde{\eta}+\frac{1}{2}\st\epsilon\alphaM \hgas\right],\\ 
    \label{eq:vdx}
    &\frac{\vdx}{C_s}=-\frac{1}{\Delta^2}\left[2\st\widetilde{\eta}+(\epsilon+1)\alphaM \hgas\right],\\ 
    \label{eq:vdy}
    &\frac{\vdy}{C_s}=-\frac{1}{\Delta^2}\left[(\epsilon+1)\widetilde{\eta}-\frac{1}{2}\st\alphaM \hgas\right], 
\end{align}
where $\Delta^2=(1+\epsilon)^2+\st^2$. The solution of these equations include effects of both pressure gradient and magnetic torque. For small grains ($\st\ll 1$), radial drift is mainly induced by the pressure gradient, while azimuthal drift is induced by magnetic torque \citepalias{Lin22}.

\subsection{AdSI in the absence of a pressure gradient}\label{linear}


\citetalias{Lin22} showed that the above accreting, dusty disk is subject to the AdSI, distinct from the classical SI. In linear theory, the AdSI can be isolated by considering perturbations without vertical dependence and by setting $\widetilde{\eta}=0$, as both are required for the classical SI. 

For linear perturbations $\propto \exp{\left(\sigma t + k_x x \right)}$, where $\sigma=s-i\omega$ is the complex frequency ($s$ and $\omega$ being the real growth rate and oscillation frequency, respectively) and $k_x$ is the radial wavenumber, a simplified dispersion relation for the AdSI reads:  
 \begin{equation} \label{adsi_disp}
     \left(\sigma \taus+\mu_d\right)\left(\sigma \taus+\mu_d+\frac{\epsilon \mathrm{St}^2}{1+\epsilon}\right)=\frac{-2i\epsilon K_x u_y \mathrm{St}^2}{1+\epsilon},
 \end{equation}
 where $\mu_d=ik_xv_{\mathrm{d}x}\taus$ and $u_y=(v_{\mathrm{d}y}-v_{\mathrm{g}y})/C_s$ is a dimensionless measurement of azimuthal drift. The normalized wavenumber $K_x = k_x\Hgas$. 

\begin{figure}  
    \includegraphics[width=\columnwidth]{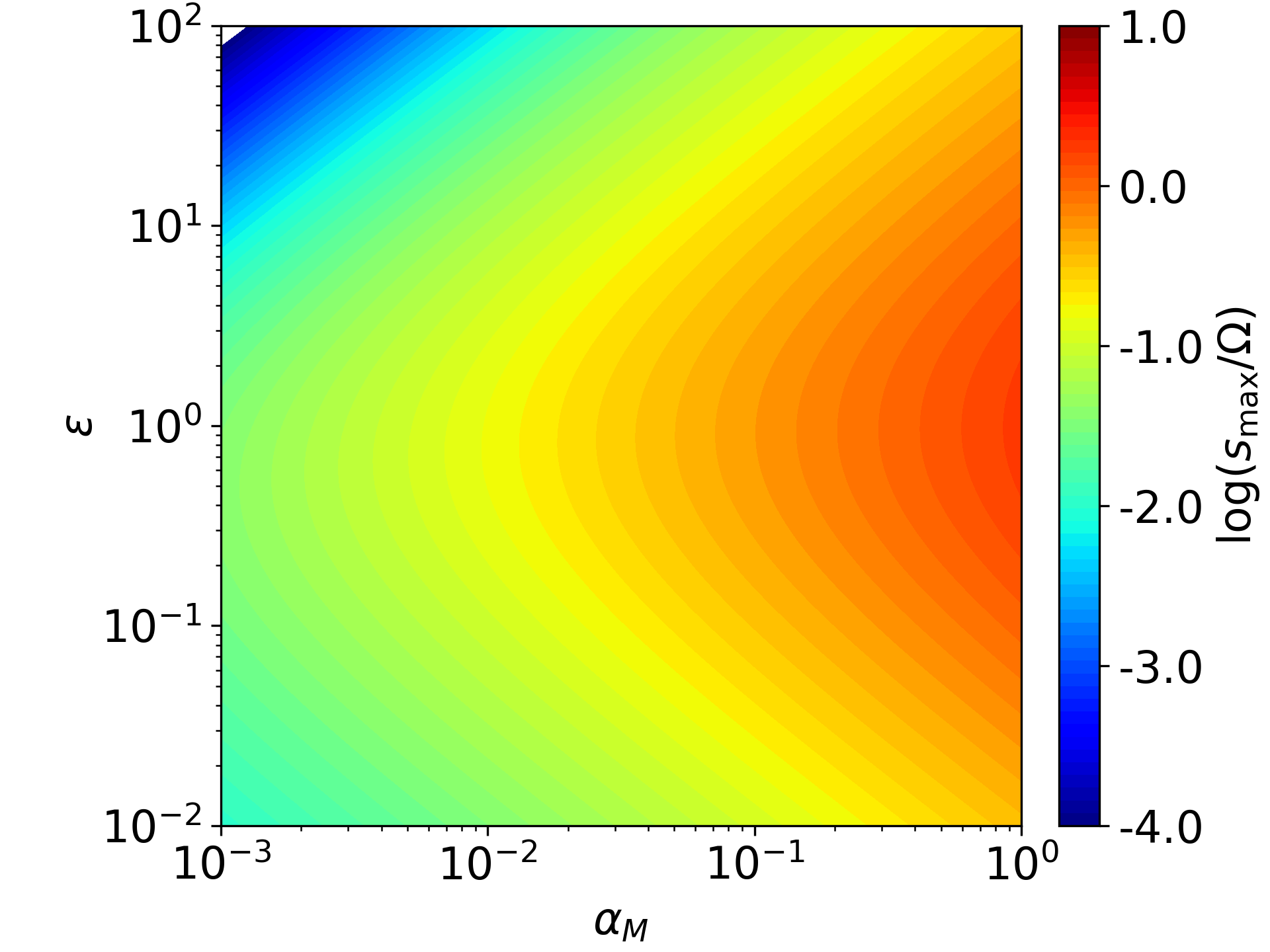}
    \caption{Linear growth rates of the AdSI as a function of $\epsilon$ and $\alphaM$ for fixed $\st=0.1$, obtained from Eq.\ref{adsi_disp} \cite[see][]{Lin22}.}
    \label{fig:am_eps_anal}
\end{figure}

In Fig. \ref{fig:am_eps_anal}, we solve Eq. \ref{adsi_disp} to obtain growth rates as a function of $\epsilon$ and  $\alphaM$ with fixed $\st=0.1$ and $K_x=5000$. The AdSI is most efficient at $\epsilon=1$, with growth rates $s\gtrsim 0.1\Omega$ for $\alphaM\gtrsim 0.01$. Indeed, we will find that the AdSI concentrates dust most effectively at intermediate $\epsilon$ between $0.1$ and $10$.


\subsection{Stabilization of the AdSI by a pressure gradient and the transition to classical SI}\label{linear_pgrad}

\begin{figure}  
    \includegraphics[width=0.48\columnwidth]{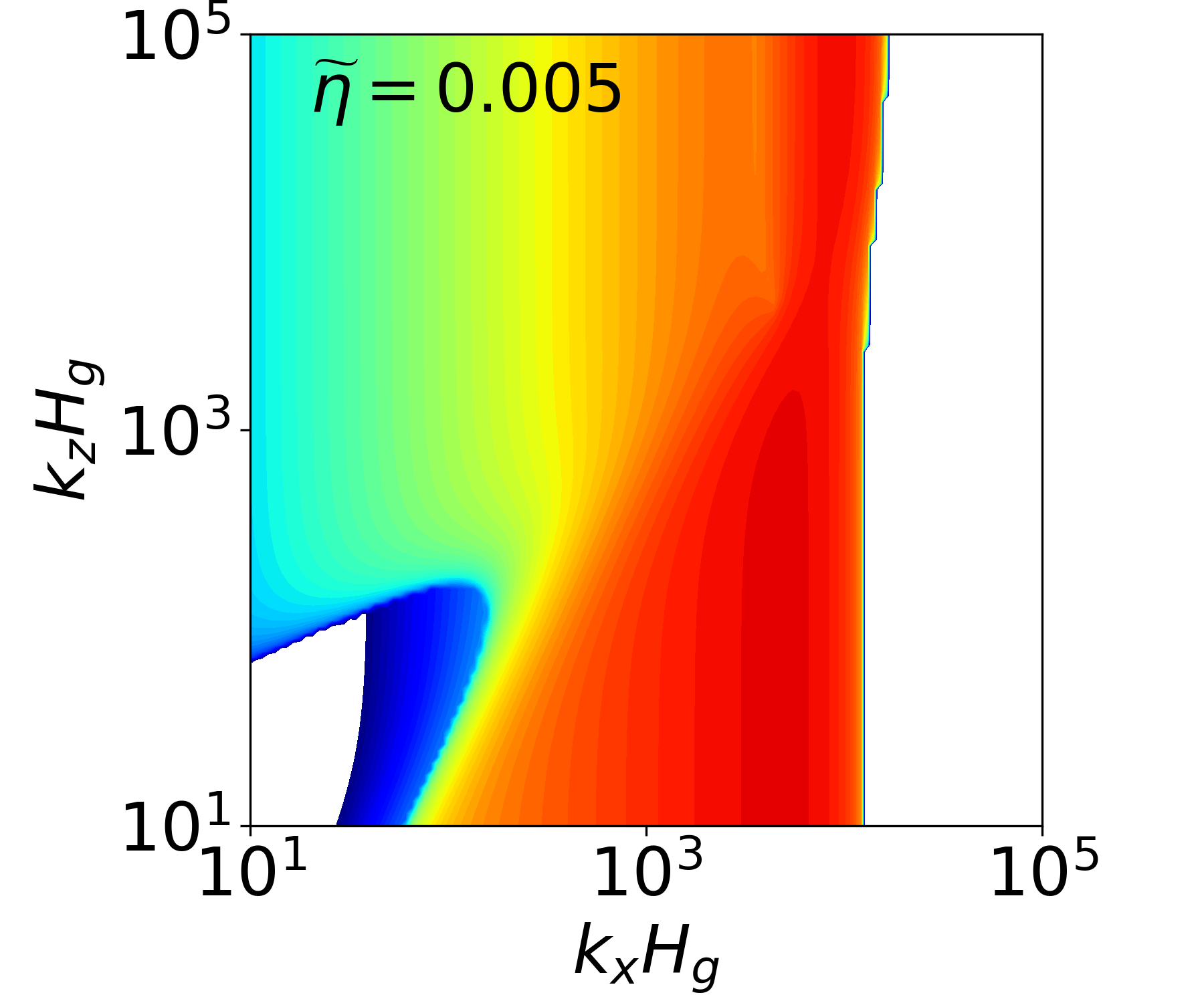}
    \includegraphics[width=0.48\columnwidth]{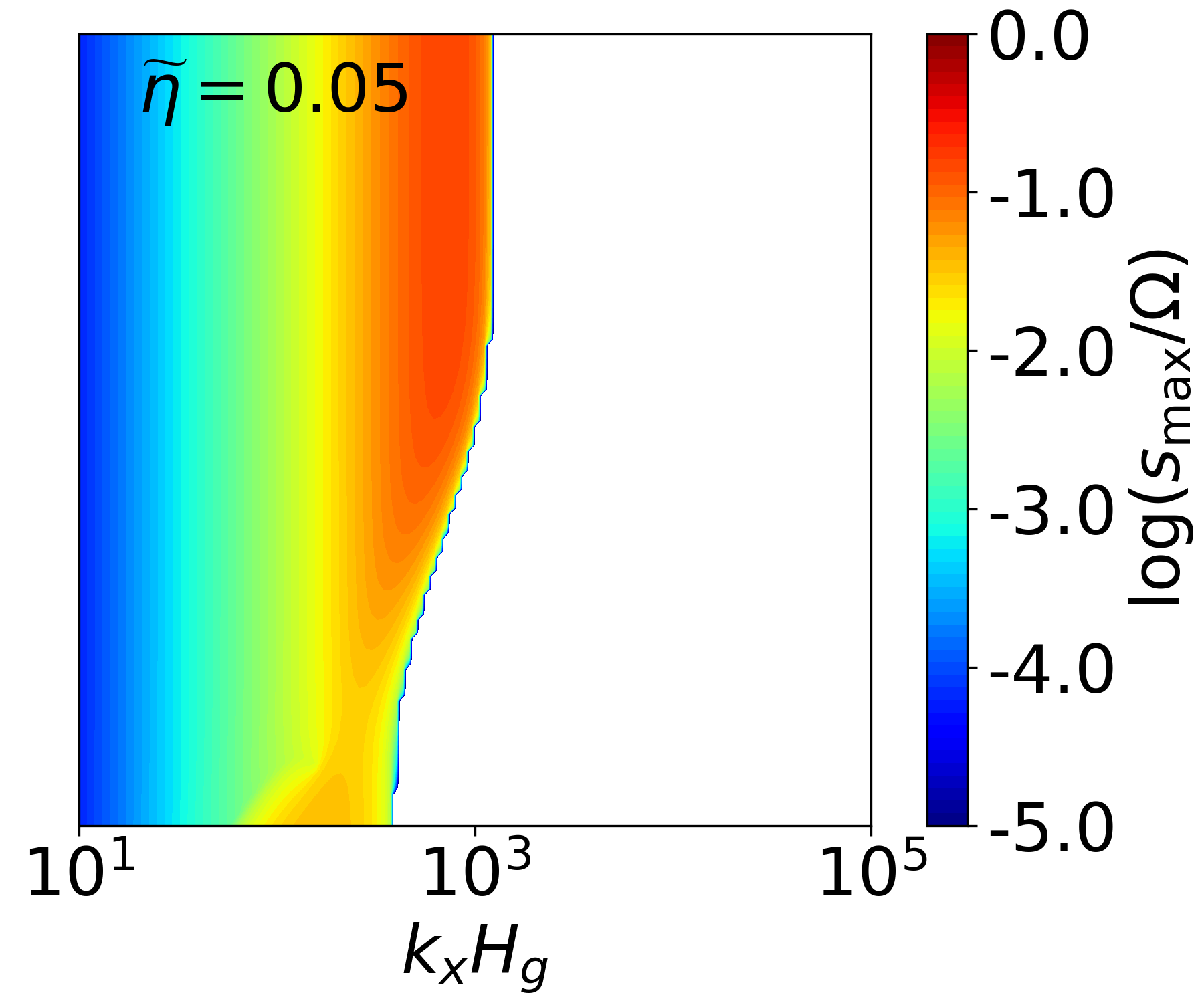}
    \caption{Growth rates of unstable modes with non-zero background pressure gradients, as a function of wavenumbers. Left: AdSI dominate case ($\widetilde{\eta}=0.005$). Right: Classical SI dominate case ($\widetilde{\eta}=0.05$). }
    \label{fig:UnstableMode}
\end{figure}

The AdSI is also distinguished from the classical SI in that the former only exists for sufficiently small $\widetilde{\eta}$. To illustrate this, we compute linear growth rates from the full stability problem described in \citetalias{Lin22} for $\widetilde{\eta}=0.005$ and $0.05$. The results are shown in Fig. \ref{fig:UnstableMode} as a function of wavenumbers. Here, we consider modes with an additional vertical dependence $\propto \exp{k_z z}$ and define the dimensionless vertical wavenumber $K_z=k_z\Hgas$. 

For $\widetilde{\eta}=0.005$, the most unstable modes are those independent of $K_z$, i.e., AdSI, extending to $K_z=0$.  Such modes are stabilized when increasing to $\widetilde{\eta}=0.05$, replaced by the classical SI. 
However, we will find that an accretion flow significantly impacts the classical SI in the nonlinear regime despite the AdSI being absent at the linear level.



\section{Numerical simulations} \label{sec:method}

We evolve the full axisymmetric shearing box equations (\ref{eqn:gascontinuous}-\ref{eqn:dustmomentum}) using the multi-fluid version of \textsc{fargo3d} \citep {Benitez-Llambay16, Benitez-Llambay19}. The code was further modified by \citetalias{Hsu22} by removing the background shear flow from the outset and introducing the azimuthal forcing $F_{\phi}$ in the gas momentum equation.



Our simulations are three-dimensional (3D) but axisymmetric. To this end, we set the $y$ grid to a single cell. We choose a computational domain of $x\in[-0.1,0.1]\Hgas$ and $z\in[-0.0125,0.0125]\Hgas$ and a resolution of $N_x\times N_z = 2048\times 256$, or 5120 cells per $\Hgas$. The box size is a compromise between convergence and the cost of running a parameter study (see Appendix \ref{sec:boxsize} for further discussions).  
We apply strictly periodic boundary conditions in $x$ and $z$ and run our simulation to $T=150P$, where $P\equiv 2\pi/\OmK$ is the orbital period, which is three times longer than \citetalias{Hsu22}. 

We adopt computational units such that $R=\Omega=1$, and set $C_\mathrm{s}= \Hgas \Omega = 0.05$. Since we omit self-gravity, the density scales are arbitrary. We thus set the initial gas density $\rho_0=1$ for convenience. 

\subsection{Parameter study and analysis} \label{sec:parameter_study}
We primarily focus on the AdSI. Thus, for the most part, we set the radial pressure gradient $\widetilde{\eta}=0$ to eliminate the classic SI. We then vary $\epsilon$ in the range $[10^{-2}, 10^2]$ and $\alphaM$ in the range $[10^{-3},1]$. Runs with $\epsilon=100$ are included for completeness, but in a realistic disk, it may immediately undergo gravitational collapse. We are especially interested in low dust-to-gas ratios with $\epsilon<1$. In this limit, the classical SI, even if it operates,  saturates at a relatively low amplitude without strong clumping \citep{Johansen07}. However, we will find that the AdSI still yields noticeable dust concentrations in dust-poor disks. 

We nominally fix $\st=0.1$ but explore $\st=10^{-2}$ and $\st=10^{-3}$. 
We present some simulations with $\widetilde{\eta}>0$ to directly compare the classic SI with the AdSI and study how the classic SI is affected by accretion flows. 

\subsubsection{Turbulence and diffusion coefficients}

For result analyses, we follow \cite{Johansen07} and define the averaged dimensionless orbital angular momentum flux
\begin{equation}
    \alphass= \overline{\left\langle\frac{\rhog \vgx \vgy}{\rho_0 C_s^2}\right\rangle}, \label{eqn:alpha_ss}
\end{equation}
where $\langle\cdot\rangle$ denotes an average over the $x$-$z$ plane and $\overline{\cdot}$ denotes a time average from $t=120P$ to $t=150P$.

We also measure the bulk gas diffusion coefficient $D_{\mathrm{g},i}$ to quantify the turbulence strength associated with the $i^\mathrm{th}$ direction. Here, we follow \cite{Yang18} and define the dimensionless diffusion coefficient associated with the $i^\mathrm{th}$ gas velocity component: 
\begin{equation}
    \alpha_{\mathrm{g},i}= \frac{D_{\mathrm{g},i}}{C_s \Hgas} \cong \left(\frac{\overline{\delta v_{\mathrm{g},i}}}{C_s}\right)^2 \Omega t_{\mathrm{corr},i},
    \label{eqn:alpha_g}
\end{equation}
where $\delta v_{\mathrm{g},i}=\sqrt{\left\langle v_{\mathrm{g},i}^2\right\rangle-\left\langle v_{\mathrm{g},i}\right\rangle^2}$ is the horizontally-averaged velocity dispersion and $t_{\mathrm{corr},i}$ is the correlation time in each direction. To compute $t_{\mathrm{corr},i}$, we follow \citetalias{Hsu22} and define the correlation time as the half-life of the autocorrelation function of the velocity field \citep[see also][]{Yang18}. 


\subsubsection{Conditions for strong clumping and gravitational collapse} \label{strong_clumping_def}

We follow \citetalias{Hsu22} and define strong clumping as filament formation with dust-to-gas ratios $\epsilon \geq 100$. Gravitational collapse is expected if the corresponding dust density, $\rhod$, exceeds the Roche density, $\rho_\mathrm{R}=9\Omega^2/4\pi G$ \citep{Klahr20}. This, in turn, depends on the physical disk model and, hence, the distance from the star. 

First, we use the Toomre parameter $Q_\mathrm{T} \equiv  C_s\Omega/\pi G \Sigma_{\mathrm{g}}$, where $\Sigma_{\mathrm{g}}=\sqrt{2\pi}\Hgas\rhog$ is the gas surface density, to write $\rho_\mathrm{R} \simeq 5.6 Q_\mathrm{T}\rhog$. Then the gravitational collapse condition, $\rhod>\rho_\mathrm{R}$, translates to $\epsilon \gtrsim 5.6 Q_\mathrm{T}$. This equals our clumping condition ($\epsilon \geq 100$) if $Q_\mathrm{T} \lesssim 18$. 

A given value of $Q_\mathrm{T}$ can be mapped to a physical distance from the star for a given disk model. To this end, we again consider the MMSN with $\Sigma_{\mathrm{g}}=2200\left( \frac{R}{\mathrm{AU}}\right)^{-3/2} \mathrm{g}\ \mathrm{cm}^{-2}$ and $H_{\mathrm{g}}=0.022R\left( \frac{R}{\mathrm{AU}}\right)^{2/7}$ \citep{Chiang10}. 
Combining these with the definition of $Q_\mathrm{T}$ and assuming a solar-mass central star, we have 
\begin{align}
    Q_\mathrm{T} \simeq 28 \left( \frac{R}{\mathrm{AU}}\right)^{-3/14.}
\end{align}
Inverting this gives $R=8\mathrm{ AU}$ for $Q_\mathrm{T}=18$. 

Thus, our clumping condition is also the criterion for gravitational collapse at $R=8\mathrm{AU}$ in the MMSN. Moving the shearing box inwards increases $Q_\mathrm{T}$ and $\rho_R/\rhog$, implying our clumping condition is no longer sufficient for collapse. Conversely, moving the box eases the collapse condition.

To anchor our results below, we will take our shearing box to be located at $R=8\mathrm{AU}$ in an MMSN-like disk so that strong clumping is expected to trigger gravitational collapse. 


\section{results} \label{sec:result}
We now present our simulation results. We first discuss simulations with varying $\epsilon$ and $\alphaM$. 
We categorize simulation outcomes as `Clumping' if $\rho_\mathrm{d,max}>\rho_\mathrm{R}$ as described in the previous section. `Turbulent' systems have reached a quasi-steady state, but $\rho_\mathrm{d,max}<\rho_\mathrm{R}$. `Unsaturated' runs have not yet converged to a quasi-steady state within the simulation time. We label runs that do not exhibit appreciable instability and remain close to their initial state as `Stable.' We are interested in clumping caused by the AdSI or SI.  Thus, for cases with an initial $\epsilon=100$, we impose an additional requirement of filament formation to be considered clumping. 

We discuss cases with different Stokes numbers in \S\ref{sec:vary_St} and compare AdSI-dominant cases and classical SI-dominant cases in \S\ref{sec:varyeta}.

\subsection{Varying $\epsilon$ and $\alphaM$, fixed $\mathrm{St}=0.1$} \label{sec:vary_eps_alpha}
\begin{figure}  
    \includegraphics[width=\columnwidth]{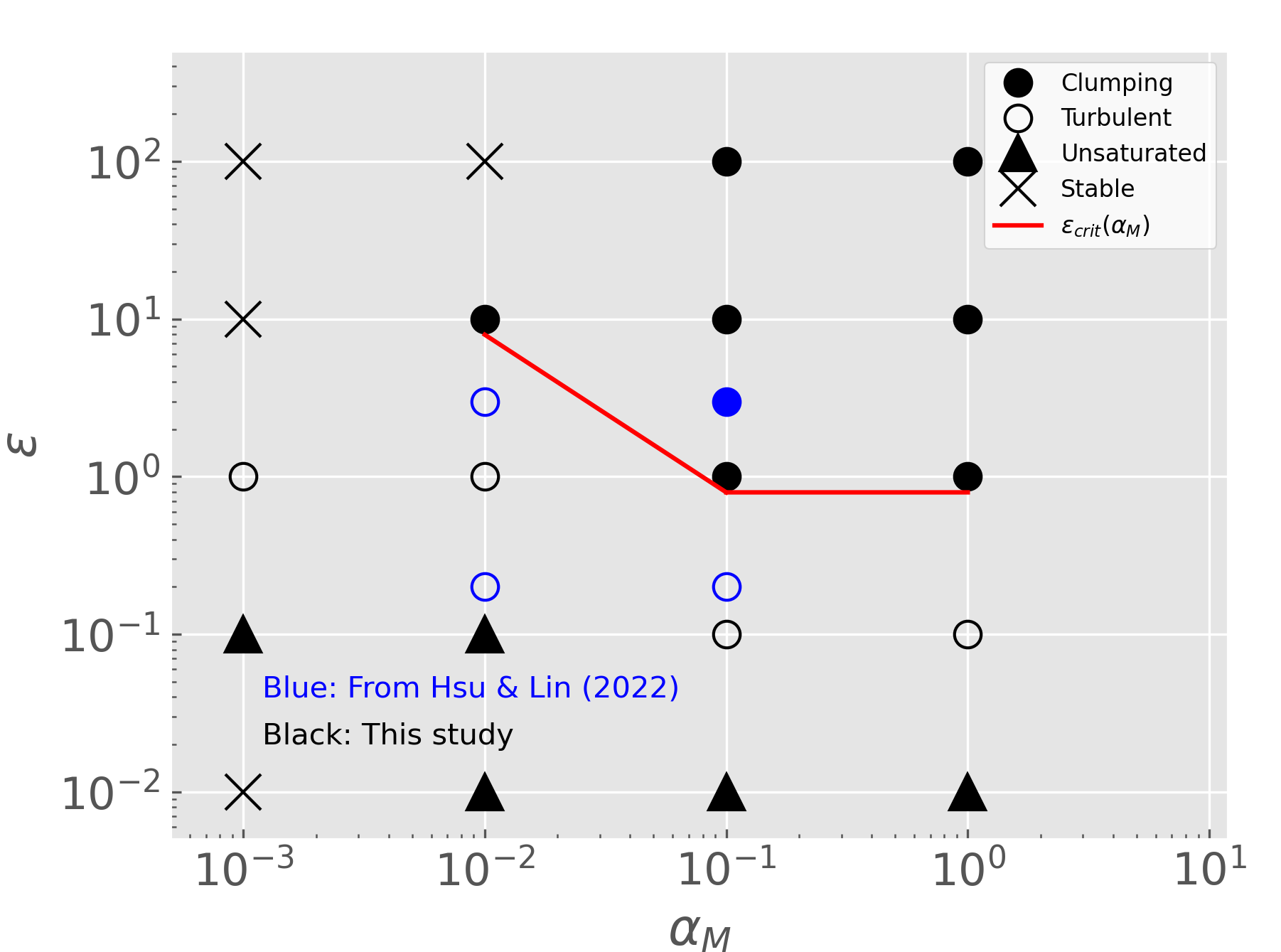}
    \caption{Overview of the end state with different initial $\alphaM$ and $\epsilon$ in our simulations. The red curve delineates the critical dust-to-gas ratio $\epsilon_{\mathrm{crit}}(\alphaM)$ as the function of $\alphaM$. Cases above this curve meet the clumping condition. We also include the relevant simulations from \citetalias{Hsu22} (in blue symbols) for comparison. }
    \label{fig:overview}
\end{figure}

We first consider different combinations of $\epsilon=\{10^{-2},10^{-1},1,10,10^2\}$ and $\alphaM=\{10^{-3},10^{-2},10^{-1},1\}$. Fig. \ref{fig:overview} gives an overview of the end states of these runs. We fit the critical dust-to-gas ratio $\epsilon_{\mathrm{crit}}$, above which strong clumping is achieved. We find $\epsilon_{\mathrm{crit}}\simeq 1$ for $\alphaM\gtrsim 0.1$, but for $10^{-2}\lesssim \alphaM  \lesssim 10^{-1}$ we find  $\epsilon_{\mathrm{crit}}\sim\alphaM^{-1}$. Thus, for $\alphaM\lesssim 10^{-2}$, the required $\epsilon_{\mathrm{crit}}$ already exceeds the clumping condition. 

\subsubsection{Dust density fluctuations and growth rate}

\begin{figure}  
    \includegraphics[width=\columnwidth]{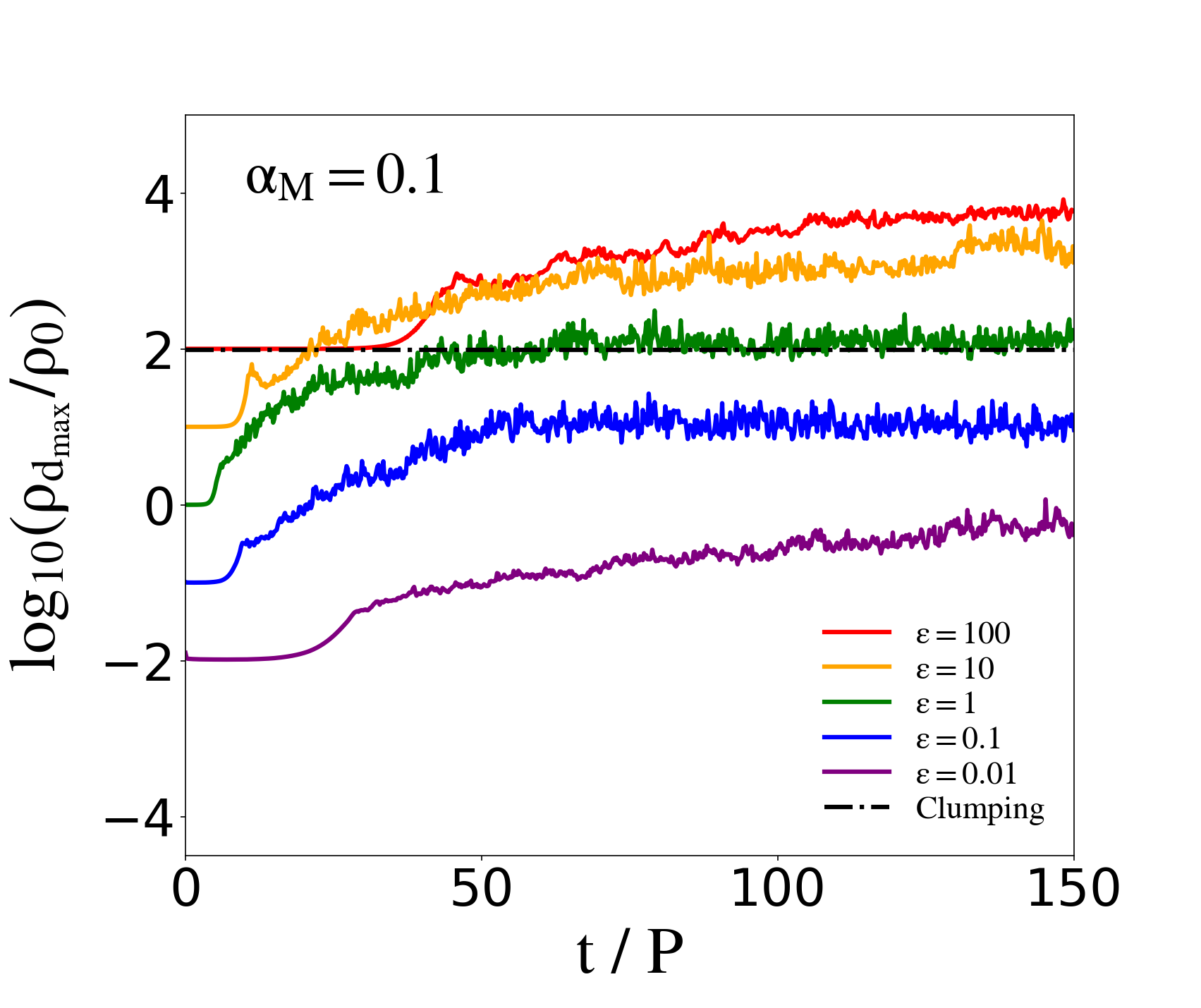}
    \centering
    \caption{Evolution of dust density perturbation of fiducial case. Panels are corresponding to different initial $\alphaM$. We use different color to indicate different initial $\epsilon$, and we also show the clumping condition in the plot (dashed-dotted line).}
    \label{fig:rhodmax_fiducial}
\end{figure}

\begin{figure*}  
    \includegraphics[width=\textwidth]{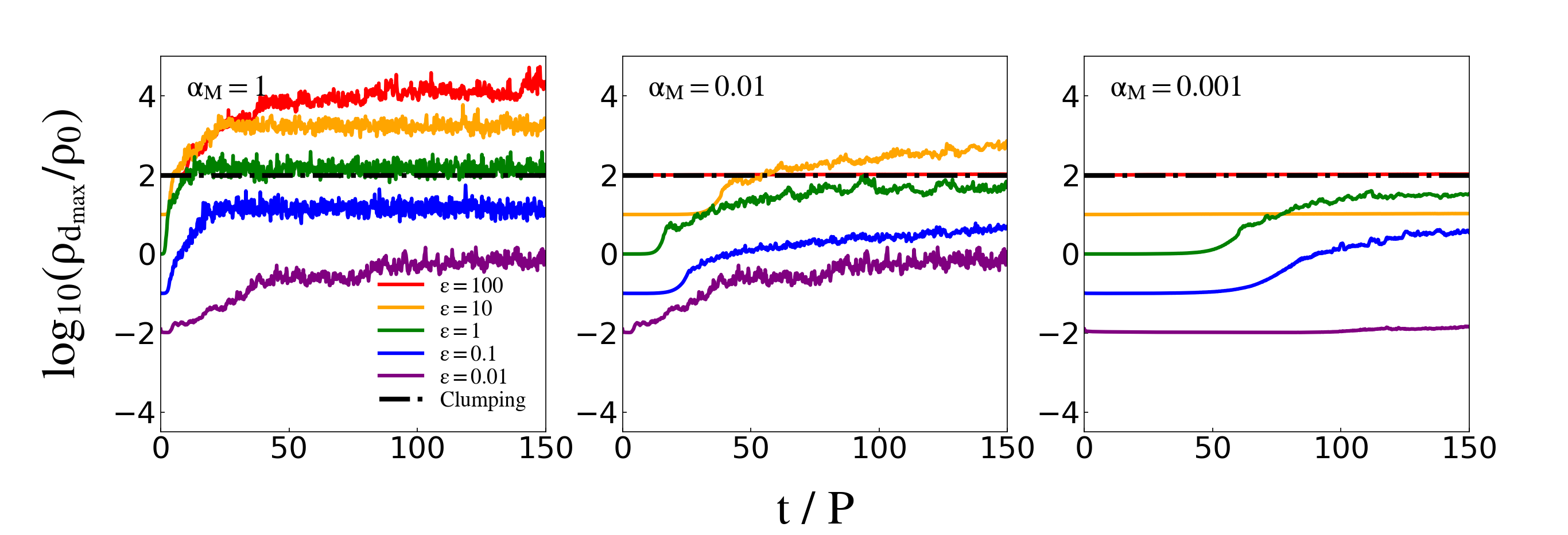}
    \centering
    \caption{Same as Fig. \ref{fig:rhodmax_fiducial} but with $\alphaM$ equals to $1,0.01,$ and $0.001$ from left to right panels. Note that for the red line in the middle panel, the AdSI saturates at negligible amplitudes and thus appears stable. This is the same as the red and yellow lines in the right panel.}
    \label{fig:rhodmax_varyam}
\end{figure*}

 We compare the maximum dust density evolution in Fig. \ref{fig:rhodmax_fiducial} and Fig. \ref{fig:rhodmax_varyam}. We normalize it by the initial gas density, but since the gas density has negligible evolution, the plot is equivalent to the maximum dust-to-gas ratio. 
 
 We start from our fiducial runs with $\alphaM=0.1$ in Fig. \ref{fig:rhodmax_fiducial}. For $\epsilon=1$ (green), the system grows rapidly in the first 10 orbits with a growth rate $s\cong 0.33 \Omega$, and then for the following $40-50$ orbits, $\rhod$ grows gentler with $s\cong 0.034 \Omega$ to reach a quasi-steady state. Here, we measure growth rates using Fig. \ref{fig:deltarhod} in Appendix \ref{sec:density_perturbation}. The final dust density meets the clumping condition, shown in the figure's dotted line. This run evolves similarly to the `E3eta0am01' case in \citetalias{Hsu22}, which adopted $\epsilon=3$. 

Decreasing the initial $\epsilon$ to $0.1$ (blue), we find the maximum dust density also drops by an order of magnitude magnitude. We observe a similar result when increasing $\epsilon$ from unity to 10 (yellow), where the maximum dust density increases by an order of magnitude.  However, increasing to $\epsilon =100$, $\rho_\mathrm{d,max}$ only becomes a few times larger than that attained by the $\epsilon =10$ case. The reason may be that the AdSI weakens in extremely dust-poor or dust-rich environments (see \S\ref{linear}). Notice both $\epsilon=100$ and $\epsilon=0.01$ runs have only just reached a quasi-steady state, while the more unstable cases in between reach a steady state well within the simulation timescale.

Fig. \ref{fig:rhodmax_varyam} show results for increasing and decreasing $\alphaM$. For a strong background accretion flow with $\alphaM=1$ (left panel) and $\epsilon=0.1,\, 1 $ and $10$, we find it takes only $\sim 40$ orbits to reach a steady state, which is roughly half the time compared to those with $\alphaM=0.1$. However, the maximum dust density attained is similar to $\alphaM=0.1$, except for $\epsilon=100$. 

For a weaker accretion flow with $\alphaM=0.01$ (middle panel), the clumping condition is only exceeded for $\epsilon=10$. Nevertheless, we still find significant dust density enhancements for $\epsilon=1$, $0.1$ $0.01$ for $\alphaM=0.1$; see below. Interestingly, for the weakest accretion flow with $\alphaM=0.001$ (right panel), only $\epsilon=0.1$ and $1$ yield appreciable growth, while smaller and larger dust-to-gas ratios remain close to their initial states. The stabilization from decreasing $\alphaM$ from $0.01$ to $0.001$ is especially dramatic for $\epsilon=10$. This suggests that, for $\epsilon>1$, larger $\epsilon$ also requires larger $\alphaM$ to destabilize.

\begin{figure}  
    \includegraphics[width=\columnwidth]{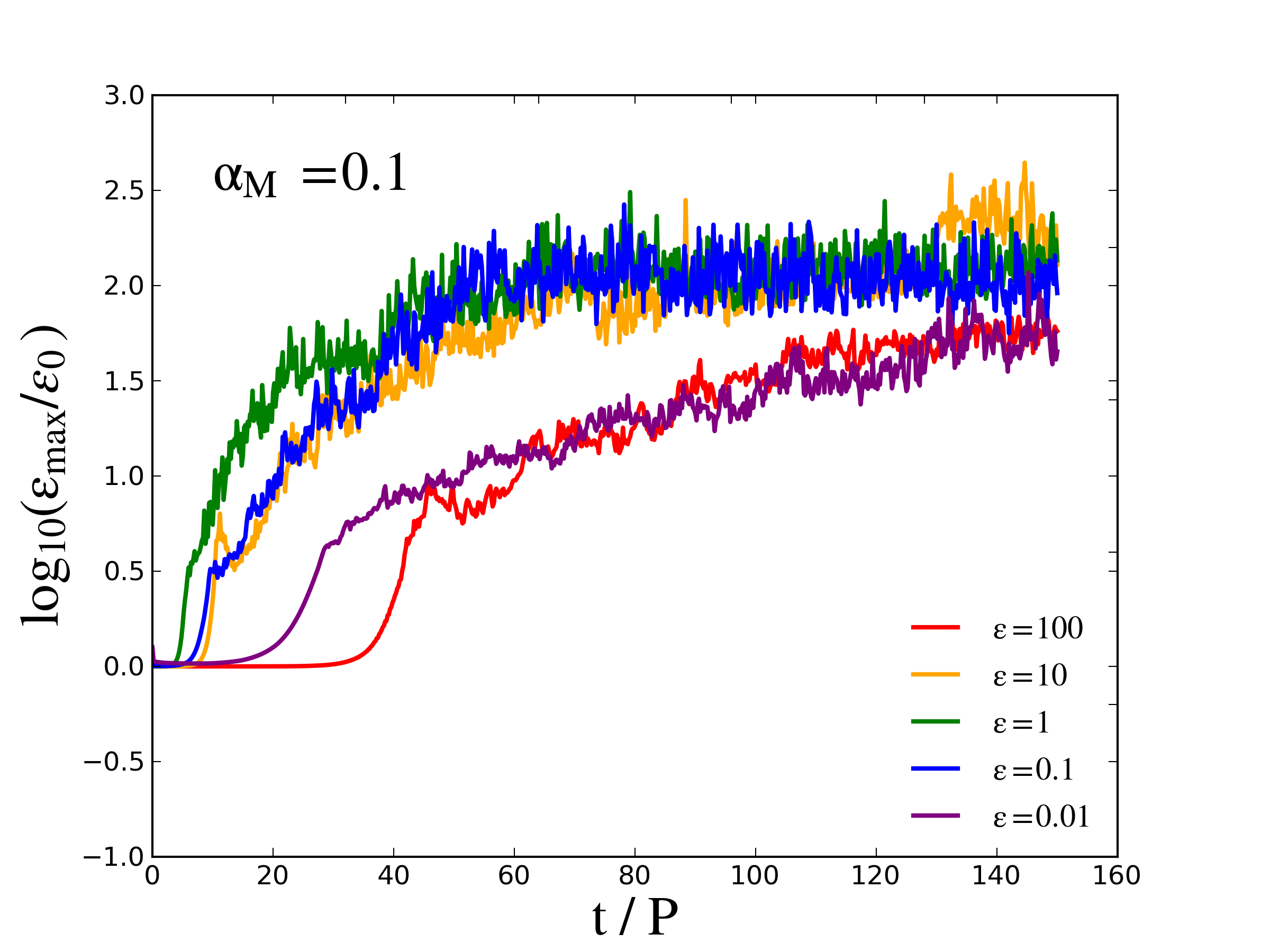}
    \caption{Evolution plot of dust density growth normalized by initial density in $\alphaM=0.1$. The color shows different initial $\epsilon$, which is the same as Fig. \ref{fig:rhodmax_fiducial}.}
    \label{fig:delta_rhod_normalized_alpha0.1}
\end{figure}

\begin{figure}  
    \includegraphics[width=\columnwidth]{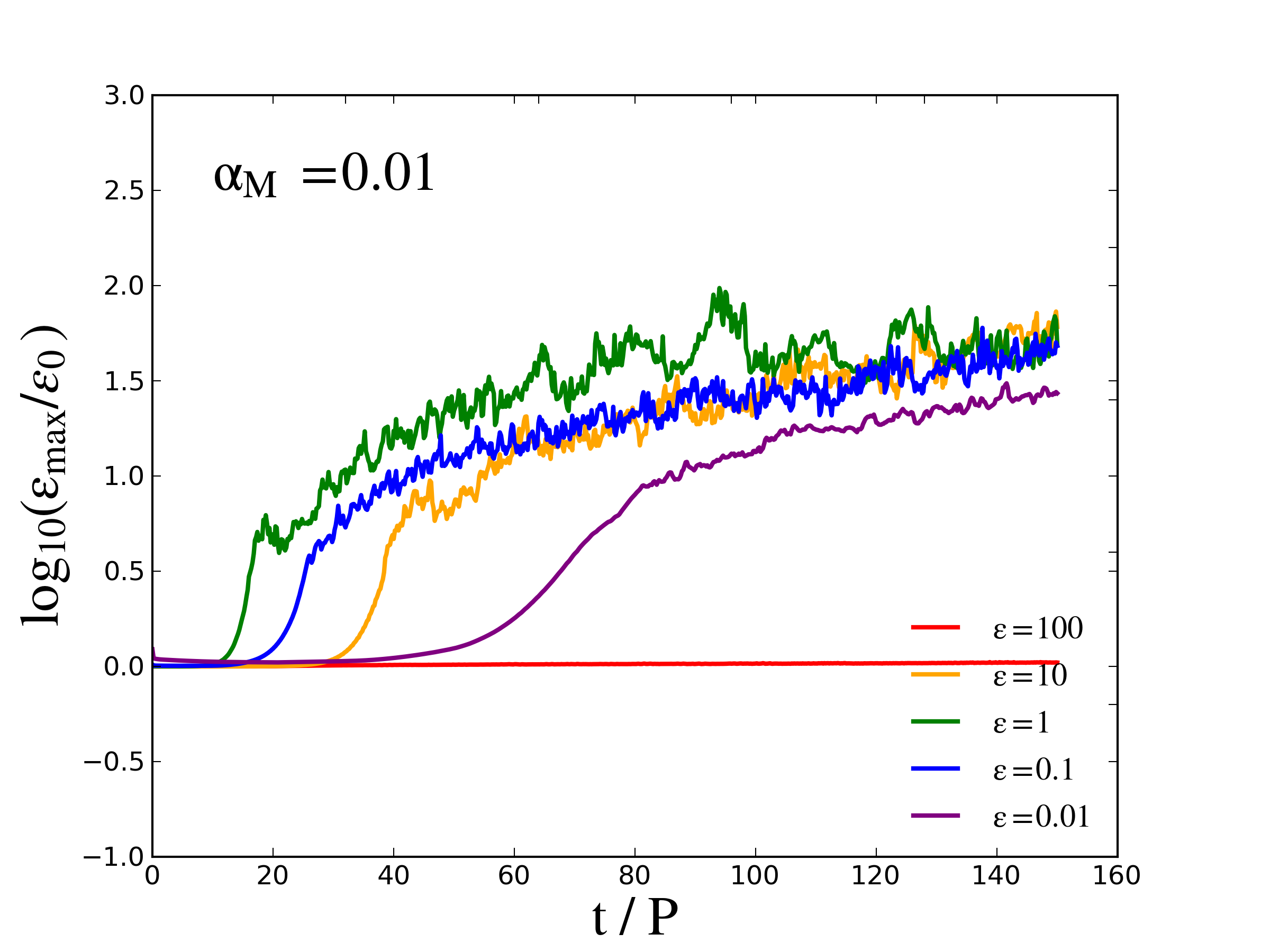}
    \caption{Same as Fig. \ref{fig:delta_rhod_normalized_alpha0.1} but with $\alphaM=0.01$.}
    \label{fig:delta_rhod_normalized_alpha0.01}
\end{figure}

We can assess the effectiveness of the AdSI in concentrating dust by normalizing the maximum dust-to-gas ratios by their initial values. These are shown in Figs. \ref{fig:delta_rhod_normalized_alpha0.1}---\ref{fig:delta_rhod_normalized_alpha0.01}.
Consider first $\alphaM=0.1$. At intermediate dust-to-gas ratios, $0.1 \lesssim \epsilon \lesssim 10$, the AdSI can increase $\epsilon$ by 100 times independently of $\epsilon$. This remains true for $\alphaM=1$, which indicates that the maximum dust density enhancement factor by the AdSI is of the order $100$. For $\epsilon=10^{\pm2}$, we find weaker concentrations with a $\sim 30$ enhancement factor.

On the other hand, Fig. \ref{fig:delta_rhod_normalized_alpha0.01} shows that even in weakly unstable disks with $\alphaM=0.01$, dust-to-gas ratios can be boosted by $\sim 30$ times, including for initial values as low as one percent. However, extremely dust-rich systems (i.e. $\epsilon=100$, red curve) remain stable. This reaffirms our previous studies that indicate the AdSI is most relevant to dust-poor regimes. 

\subsubsection{The propensity of filament merging}

\begin{figure*}  
    \includegraphics[width=\textwidth]{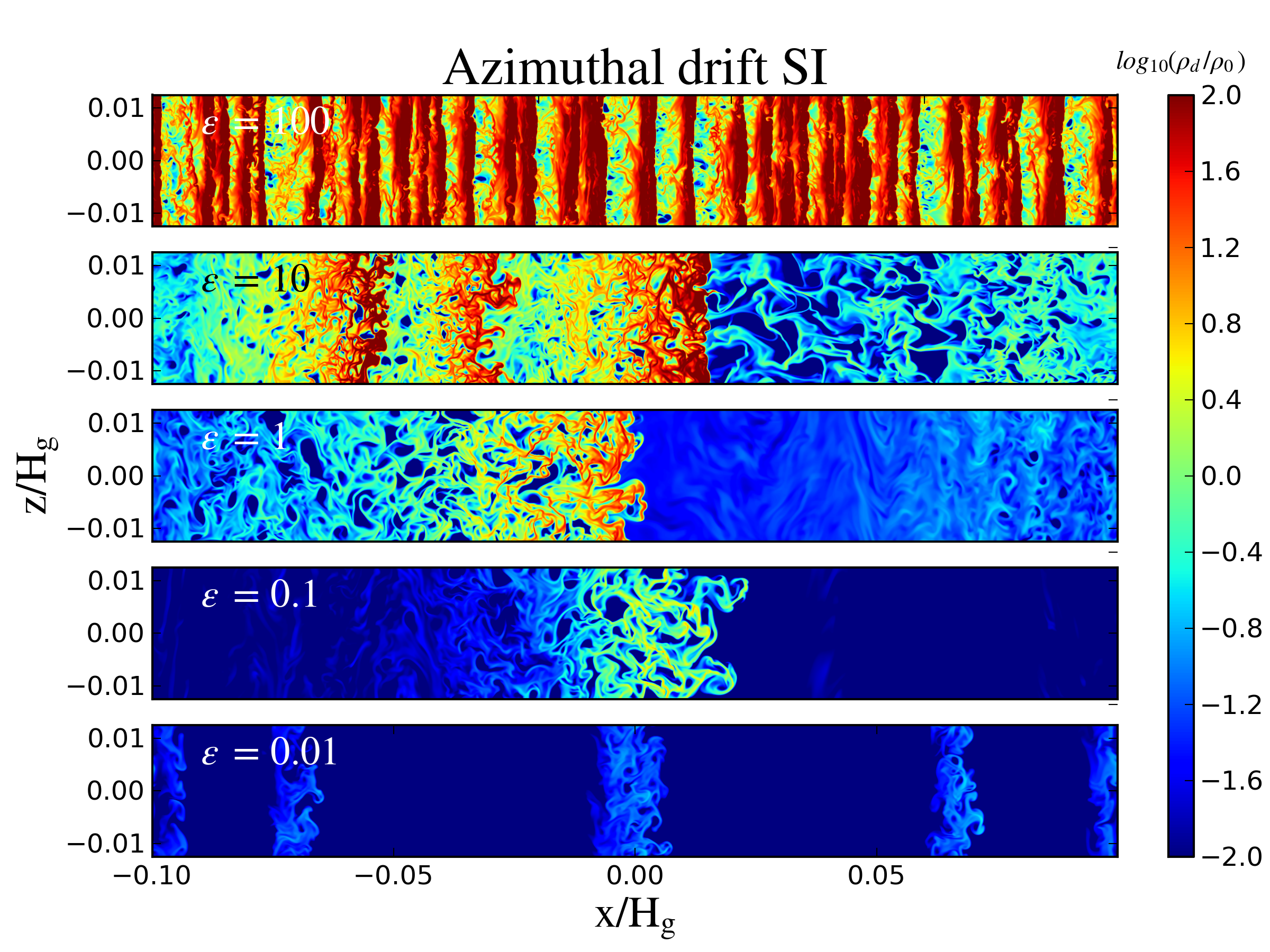}
    \caption{Snapshots of dust density of $\alphaM=0.1$. The time is set to be $t=100P$, when the disks are in a quasi-steady state}. From top to bottom panels indicate the case with $\epsilon=10^2,10^1,10^0,10^{-1}$ and $10^{-2}$ respectively. 
    
    \label{fig:density}
\end{figure*}

Fig. \ref{fig:density} shows the steady-state dust density distributions for runs with $\alphaM=0.1$ and different $\epsilon$. The AdSI primarily produces vertically elongated filaments, consistent with the linear theory showing that instability is intrinsically one-dimensional without vertical dependence \citepalias{Lin22}. Note that filaments form even in dust-poor disks with $\epsilon<1$, which is unlike the classical SI \citep[e.g.][]{Johansen07}.

As found by \citetalias{Hsu22}, the maximum dust density grows mainly through merging events between filaments. This readily occurs for $0.1 \lesssim \epsilon \lesssim 10$, the middle three panels in Fig. \ref{fig:density}. Dust filaments have little radial drift in the highly dust-rich disk with $\epsilon=100$ (top) panel. On the other hand, in the extremely dust-poor disk with $\epsilon=0.01$, the initial filaments are widely separated. As a consequence, we observe limited merging events in these extreme cases. 

\begin{figure*}  
    \includegraphics[width=\textwidth]{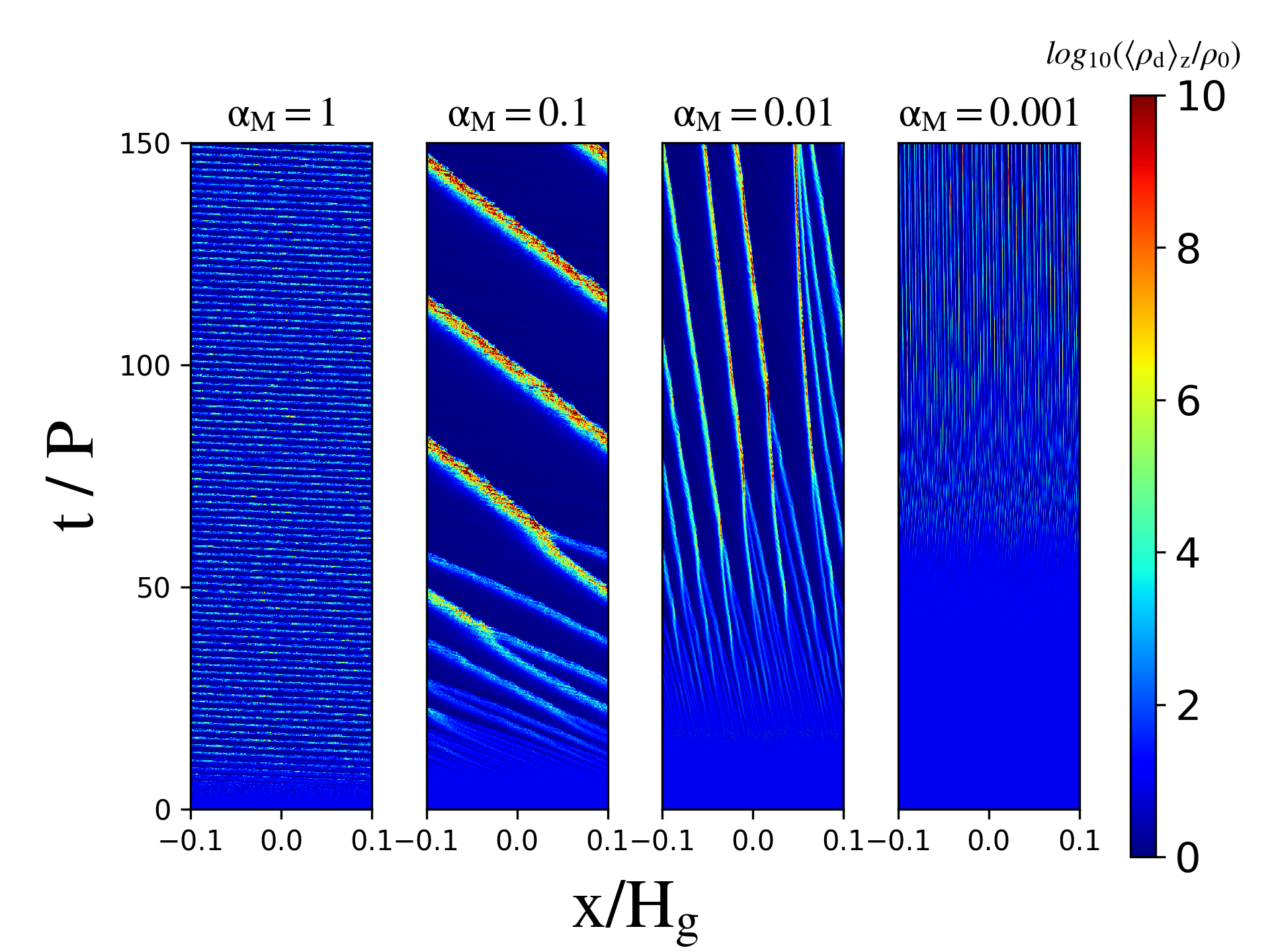}
    \caption{Space-time plot of dusty filaments with $\epsilon=1$. We compute the vertical average density at every time to show the evolution of filaments within our simulation timescale. From left to right panel shows the case with $\alphaM=10^0,10^{-1},10^{-2}$ and $10^{-3}$ respectively.}
    \label{fig:spacetime_eps1}
\end{figure*}

The relation between the maximum dust density and the propensity of filament merging can be better appreciated in space-time diagrams. This is shown in Fig. \ref{fig:spacetime_eps1} for simulations with fixed $\epsilon=1$ and varying $\alphaM$. Here, we plot the vertically-averaged dust density, which is appropriate since the AdSI has a limited vertical structure, even in its nonlinear evolution. We see that limited dust density enhancements are attained for either the largest or smallest $\alphaM$, where filaments drift inwards most or least rapidly. Instead, only for $\alphaM=0.01$ and $0.1$ do merging events readily occur and raise the dust densities. 





\subsubsection{Turbulence properties}

We now compare the strength of AdSI-driven turbulence in terms of angular momentum transport, $\alpha_\mathrm{SS}$, and gas diffusion $\alpha_{\mathrm{g},i}$, see Eqs. \ref{eqn:alpha_ss}---\ref{eqn:alpha_g} in \S\ref{sec:parameter_study}. We list the measured values for each run in Table \ref{table:turbulent}. For the small grains considered here, the corresponding particle diffusion coefficients are expected to be close to $\alpha_{\mathrm{g},i}$ as they only differ by a factor of $1+\st^2$ \citep{Youdin07b,Youdin11}. 

As found by \citetalias{Hsu22}, the AdSI is highly anisotropic with $\alpha_{\mathrm{g},y}$ is typically at least an order of magnitude larger than $\alpha_{\mathrm{g},x}$ and $\alpha_{\mathrm{g},z}$.
We find that $\alpha_{\mathrm{g},i}$ is strongly affected by $\alphaM$. For a given initial dust-to-gas ratio, say $\epsilon=0.1$, if we decrease $\alphaM$ by an order of magnitude, the diffusion coefficients also decrease by 1---2 orders of magnitude, showing that $\alpha_{\mathrm{g},i}$ is positively correlated to the accretion flow speed. On the other hand, if we fix $\alphaM$ and vary $\epsilon$, we find the value of $\alpha_{\mathrm{g},i}$ is largely unaffected to the order of magnitude. This insensitivity is unlike the classical SI, wherein diffusion weakens with increasing dust densities \citep{Schreiber18}.


The fifth column in Table \ref{table:turbulent} shows the time-averaged angular momentum transport coefficient $\alphass$. With negligible AdSI (runs Eps0.01am0.001, Eps10am0.001, Eps100am0.01, and Eps100am0.001), we have $\alphass<0$ due to the equilibrium flow. The AdSI yields a positive transport, which overcomes the background value to give $\alphass>0$ when the instability is sufficiently vigorous.
We find a similar trend to the diffusion coefficients in that $\alphass$ is nearly proportional to $\alphaM$, but insensitive to $\epsilon$. 

We find $\alphass$ is typically an order of magnitude larger than $\alpha_{\mathrm{g},x}$, i.e. the Schmidt number $\mathrm{Sc}=\alphass/\alpha_{\mathrm{g},i}\sim 10$ for most cases. This emphasizes that these coefficients represent distinct physical effects and thus have no reason to be equal.


\begin{table*}[ht]
\caption{Turbulence measurement of gas diffusion coefficient $\alpha_{\mathrm{g},i}$ in each direction, together with angular momentum flux $\alphass$. Runs are labeled by their corresponding $\epsilon$ and $\alphaM$ (e.g. Run Eps0.1am1 represent the case with $(\epsilon,\alphaM)=(0.1,1)$). }
\setlength\tabcolsep{16pt}

\begin{tabularx}{\linewidth}{ l l l l l l l l}
\hline\hline
                  
Run            & $\alpha_{\mathrm{g},x}$   & $\alpha_{\mathrm{g},y}$   & $\alpha_{\mathrm{g},z}$   & $\alphass$  & $t_{\mathrm{corr},x}$  & $t_{\mathrm{corr},y}$ & $t_{\mathrm{corr},z}$  \\
\hline
Eps0.1am1      &  7.07e-07         &  2.68e-05         &  1.49e-05         &  5.80e-05       &    0.01       &   0.11       &   0.03         \\
Eps0.1am0.1    &  1.61e-07         &  2.66e-05         &  2.86e-06         &  1.12e-06       &    0.06       &   1.30       &   0.19\\
Eps0.1am0.01   &  9.29e-09         &  1.22e-06         &  6.29e-08         &  1.69e-08       &    0.18       &   1.70       &   0.32\\
Eps0.1am0.001  &  4.57e-11         &  8.40e-09         &  3.91e-10         &  1.64e-10       &    0.21       &   1.30       &   0.36\\
Eps0.01am1     &  9.54e-07         &  2.90e-06         &  1.13e-06         &  2.84e-06       &    0.01       &   0.09       &   0.01\\
Eps0.01am0.1   &  8.27e-09         &  8.61e-07         &  4.09e-08         &  3.77e-08       &    0.01       &   0.30       &   0.05\\
Eps0.01am0.01  &  4.69e-11         &  5.80e-09         &  1.59e-10         &  3.73e-10       &    0.12       &   0.25       &   0.08\\
Eps0.01am0.001 &  2.53e-10         &  1.20e-10         &  2.55e-10         &  -5.86e-11      &    0.18       &   0.17       &   0.18\\
Eps1am1        &  4.01e-06         &  1.15e-04         &  3.76e-06         &  2.68e-04       &    0.03       &   0.24       &   0.03\\
Eps1am0.1      &  3.13e-07         &  9.46e-05         &  8.81e-06         &  9.82e-06       &    0.05       &   1.80       &   0.42\\
Eps1am0.01     &  2.43e-08         &  3.25e-06         &  4.06e-07         &  1.63e-07       &    0.13       &   1.60       &   0.55\\
Eps1am0.001    &  4.68e-10         &  4.06e-08         &  1.28e-08         &  1.40e-09       &    0.30       &   1.75       &   1.01\\
Eps10am1       &  2.50e-06         &  1.28e-03         &  1.21e-06         &  2.67e-04       &    0.03       &   1.50       &   0.03\\
Eps10am0.1     &  6.58e-07         &  1.84e-04         &  2.61e-06         &  2.83e-05       &    0.07       &   1.20       &   0.18\\
Eps10am0.01    &  8.19e-09         &  1.57e-06         &  6.63e-08         &  7.73e-08       &    0.18       &   1.25       &   0.48\\
Eps10am0.001   &  7.05e-11         &  2.15e-11         &  1.63e-11         &  -2.79e-11      &    0.18       &   0.18       &   0.19\\
Eps100am1      &  2.08e-06         &  4.35e-04         &  8.95e-07         &  2.10e-05       &    0.03       &   0.55       &   0.03\\
Eps100am0.1    &  5.64e-08         &  5.40e-06         &  6.30e-08         &  1.10e-07       &    0.10       &   0.53       &   0.14\\
Eps100am0.01   &  6.20e-11         &  1.76e-11         &  1.22e-11         &  -2.55e-11      &    0.18       &   0.17       &   0.18\\
Eps100am0.001  &  6.82e-11         &  2.08e-11         &  1.59e-11         &  -2.85e-11      &    0.18       &   0.18       &   0.19\\
\hline\hline
\end{tabularx}
\label{table:turbulent}
\end{table*}

\subsection{Varying $\mathrm{St}$, fixed $\epsilon=1$ and $\alphaM=0.1$} \label{sec:vary_St}

We next examine the AdSI for more tightly-coupled dust with $\mathrm{St}=10^{-2}$ and $\st=10^{-3}$. Using Eq. \ref{eqn:sttoparticle}, these correspond to mm and sub-mm sized dust grains, respectively. Here, we fix $(\epsilon,\alphaM) = (1, 0.1)$. 

\begin{figure}  
    \includegraphics[width=\columnwidth]{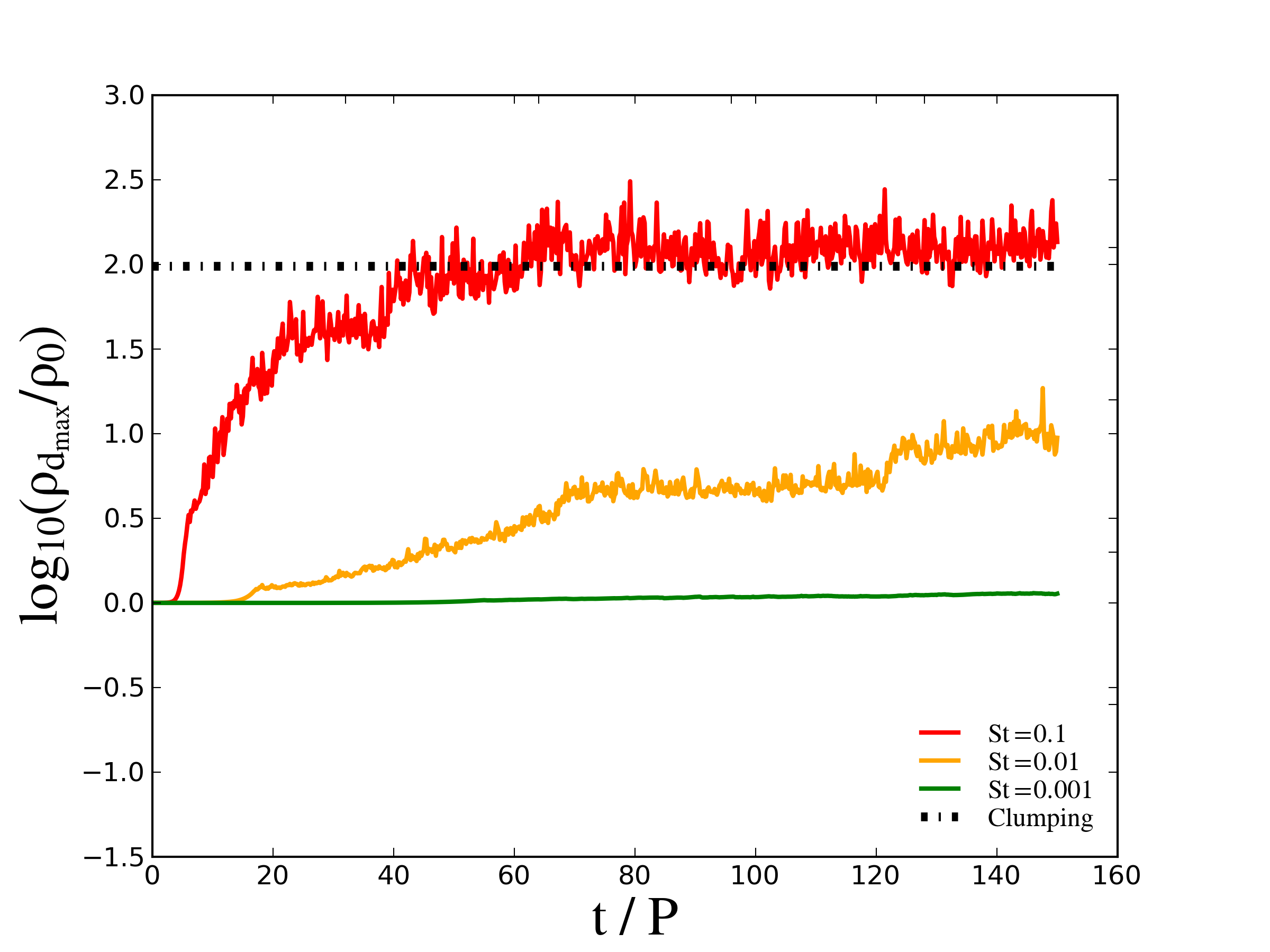}
    \caption{Evolution of dust density normalized by initial dust. Color lines indicate disks with different initial grain sizes. }
    \label{fig:rhodmax_St}
\end{figure}

Fig. \ref{fig:rhodmax_St} shows the evolution of the maximum dust densities (green and yellow curves) for these cases compared to our fiducial setup with $\st=0.1$ (red curve). Decreasing the Stokes number by a factor of 10 also decreases $\rho_\mathrm{d,max}$ by the same amount. The disk with $\mathrm{St}=10^{-3}$ is essentially stable. Since an $\alphaM$ of $O(0.1)$ is expected to be the maximum value associated with stresses driven by magnetic torques \citep{Bethune17}, the AdSI is most relevant for large grains with $\st\gtrsim 0.1$, unless the dust-to-gas ratio already exceeds unity. 

\begin{figure}  
    \includegraphics[width=\columnwidth]{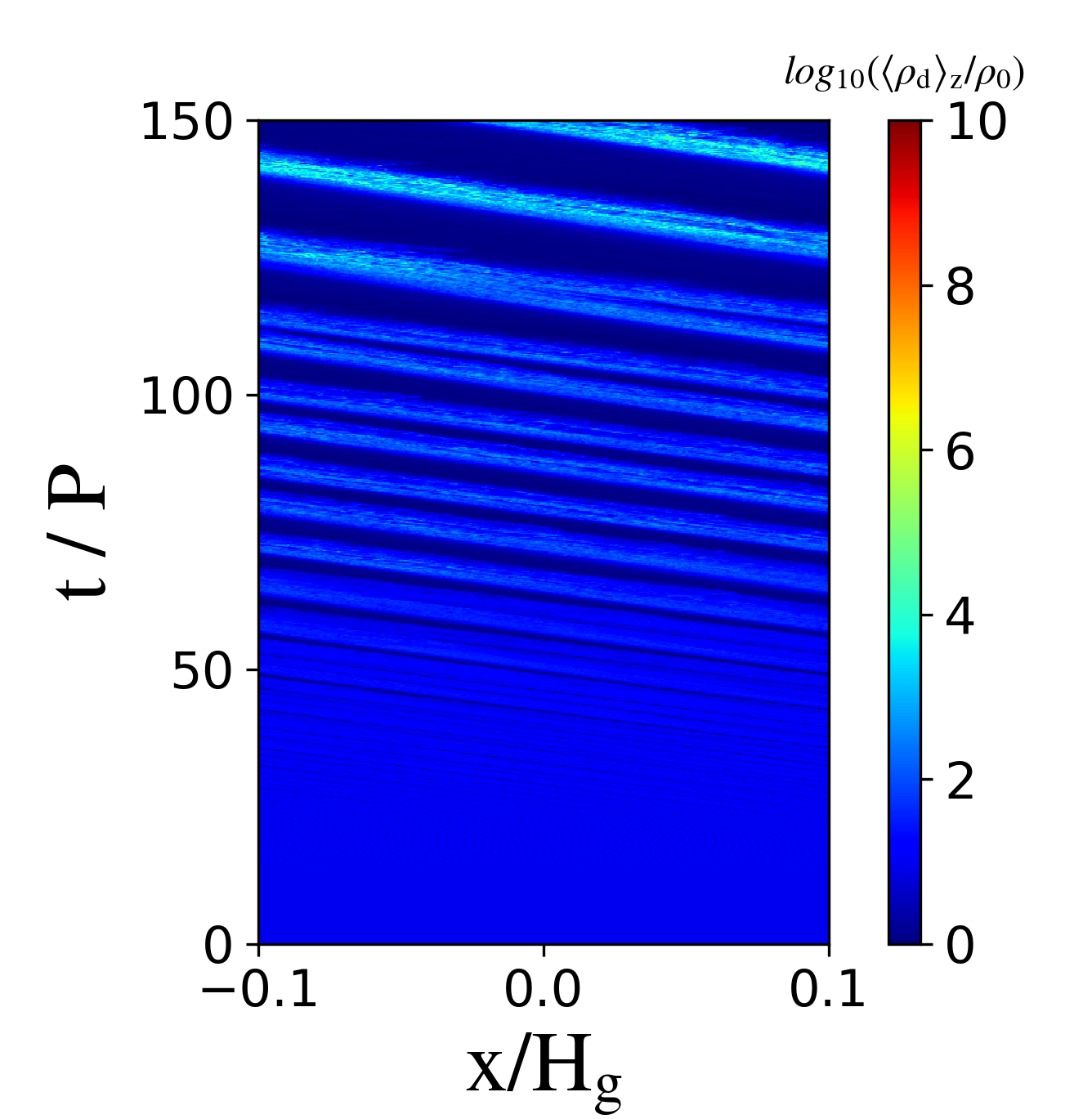}
    \caption{Similar to Fig. \ref{fig:spacetime_eps1} but for $\alphaM=0.1$ and $\mathrm{St}=0.01$.}
    \label{fig:spacetime_St}
\end{figure}

Fig. \ref{fig:spacetime_St} shows the vertically-averaged dust density evolution of the disk with $\mathrm{St}=0.01$. We can compare it with the second panel in Fig. \ref{fig:spacetime_eps1}. The filaments in the low Stokes number case have faster drift speeds (the slope is $\sim2$ times larger in Fig. \ref{fig:spacetime_St}). This difference is qualitatively consistent with the equilibrium dust drift velocity due to the applied gas torque only (i.e., Eq. \ref{eq:vgx}-\ref{eq:vdy} with $\widetilde{\eta}=0$), which shows that $v_{\mathrm{d}x}\propto (1+\epsilon)^{-1}$ for $\st\ll 1$ and is independent of $\st$. Thus, while radial drift driven by pressure gradients slows down with grain size, this effect is absent in torque-driven drifts. For $\st=0.01$, filaments attain smaller dust-to-gas ratios, which results in faster drift. 


\subsection{Comparisons with the classical SI} \label{sec:varyeta}

\begin{figure}  
    \includegraphics[width=\columnwidth]{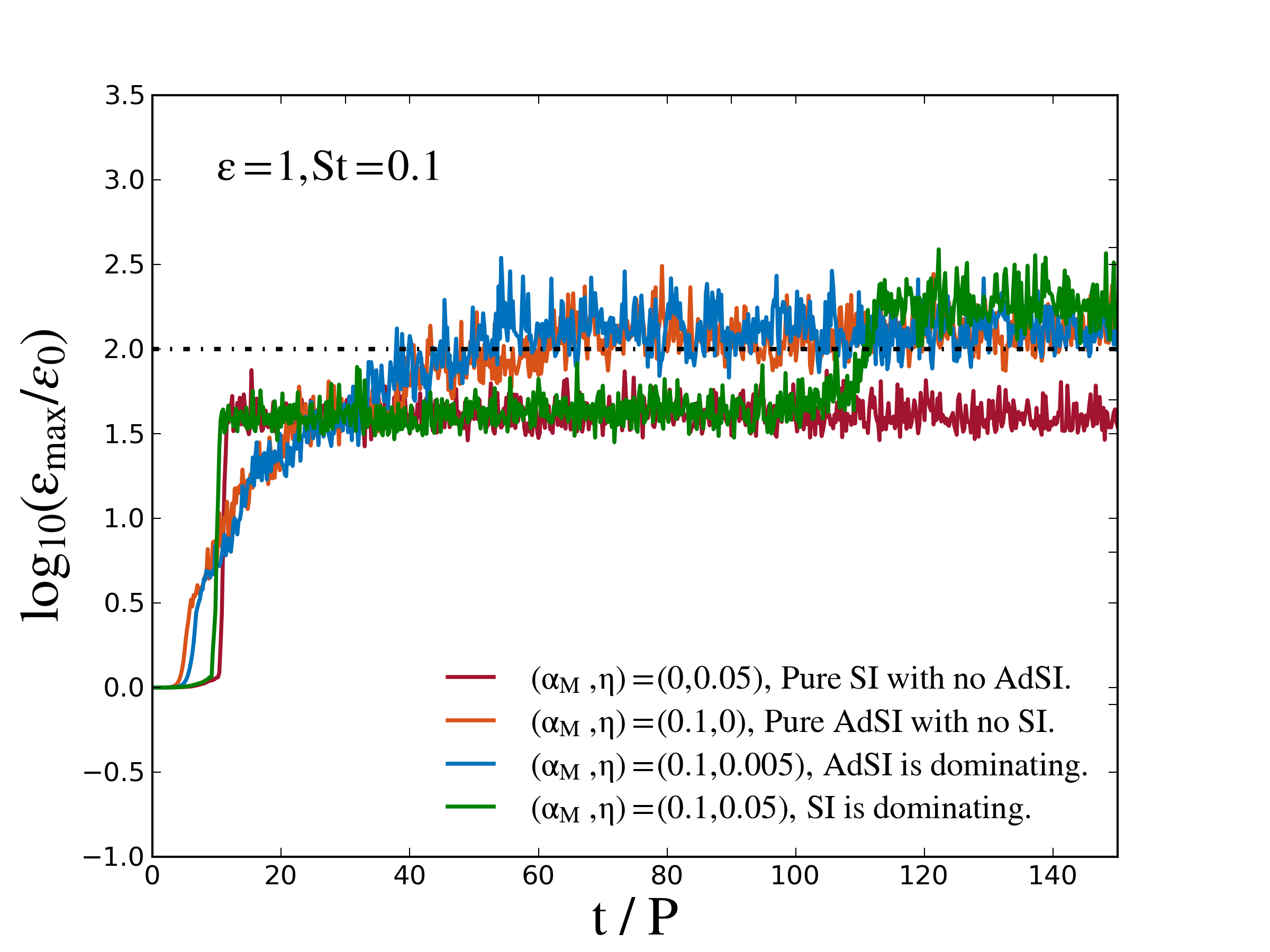}
    \caption{Similar to Fig. \ref{fig:rhodmax_fiducial} but for disk with radial pressure gradient $\widetilde{\eta}=0.05$. In three cases (dark red, orange, and dark green lines), we set the initial $\alphaM=0.1$.}
    \label{fig:classical}
\end{figure}

We now consider the effect of non-zero pressure gradients. We run a case with $\widetilde{\eta}=0.005$, for which the AdSI dominates, although the classical SI is also present. To compare AdSI with the classical SI, we run two additional simulations with $\widetilde{\eta}=0.05$: one with $\alphaM=0$ and one with $\alphaM=0.1$. We fix $\epsilon=1$ and $\st=0.1$. For $\alphaM=0$, the AdSI is strictly absent, corresponding to pure classical SI. The classical SI still dominates the $\alphaM=0.1$ case regarding linear stability, but has an accretion flow. Results are shown in Fig. \ref{fig:classical}. Our fiducial case with $\widetilde{\eta}=0$ is re-plotted (as the orange curve) for ease of reference. 

As the figure shows, with $\alphaM=0.1$, dust density growth is nearly at the same level ($\sim 100$ times) for both $\widetilde{\eta}=0$ (orange) and $\widetilde{\eta}=0.005$ (blue). This indicates that for sufficiently small $\widetilde{\eta}$, e.g. within a pressure bump in a global disk, the system behaves similarly to vanishing pressure gradients. The presence of the classical SI does not appear to further concentrate dust, although this could be due to the fact that for small $\widetilde{\eta}$, the classical SI appears at large vertical wavenumber that are not well-captured with finite resolutions. 



\begin{figure}  
    \includegraphics[width=\columnwidth]{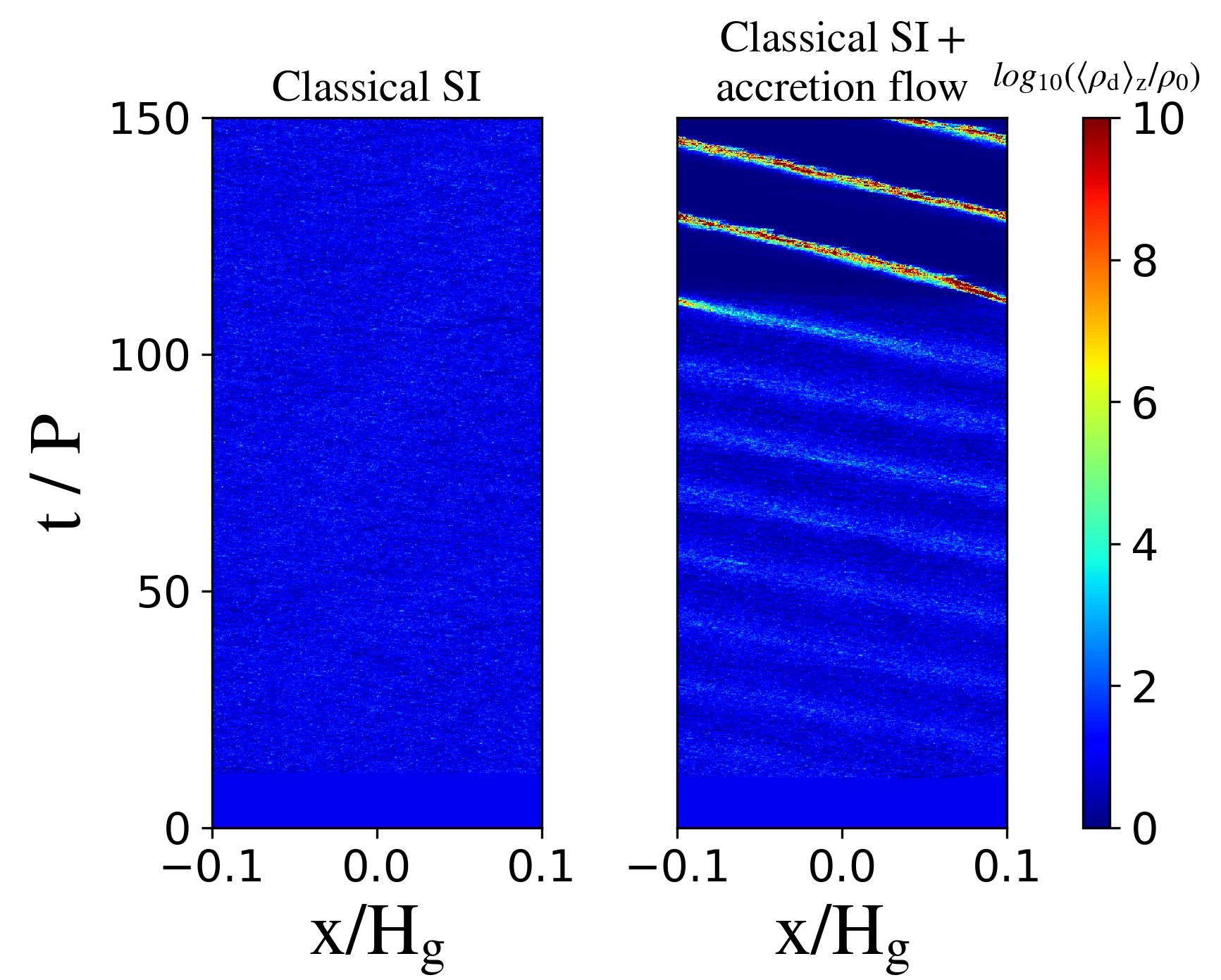}
    \caption{Similar to Fig. \ref{fig:spacetime_eps1} but with $\widetilde{\eta}=0.05$ and varying $\alphaM$. Left: Pure classical SI and $\alphaM=0$ (dark red line in Fig. \ref{fig:classical}). Right: classical SI with some background accretion flow $\alphaM=0.1$ (dark green line in Fig. \ref{fig:classical}).  }
    \label{fig:pureSI_comparison}
\end{figure}

We find the pure SI case ($\alphaM=0$, dark red) initially grows marginally faster than the pure AdSI run (orange), but the former saturates at lower amplitudes ($\epsilon_\mathrm{max}\sim 30$), which fails to meet the critical dust-to-gas ratio needed for gravitational collapse. We also find no filaments form in this case. This is shown in the left panel of the space-time plots in Fig. \ref{fig:pureSI_comparison}. In contrast, when $\alphaM=0.1$ (right panel), we do find filament formation as soon as the system saturates at $\sim 20P$. The filament rapidly intensifies around $t=110P$ and exceeds the clumping criterion. This shows that, although the AdSI is absent at the linear level, accretion flows promote clumping in the nonlinear regime even for underlying turbulence driven by the classical SI.

\section{Discussion} \label{sec:disscussion}

\subsection{Implications for planetesimal formation}

Our simulations show that the AdSI can raise local dust-to-gas ratios, $\epsilon$, by a factor of $\sim 10^2$ provided the gas accretion flow is sufficiently strong ($\alphaM\gtrsim 0.1$) and grains are relatively large ($\st\sim 0.1$).
This can trigger a gravitational collapse in the outer parts of an MMSN-like disk ($R\gtrsim 8\mathrm{AU}$), provided $\epsilon\gtrsim 1$ initially. Weaker accretion flows with $\alphaM=0.01$ achieve concentration factors $\sim 30$. Gravitational collapse at the same radii is expected to require $\epsilon\gtrsim 3$. Alternatively, if $\epsilon\simeq 1$, then gravitational collapse requires a Toomre $Q_\mathrm{T}\lesssim 5.4$ (see \S\ref{strong_clumping_def}).

Unlike the classical SI, the AdSI can operate with a vanishing radial pressure gradient \citepalias{Lin22}. This suggests the AdSI to be most applicable at or near pressure bumps. For a Gaussian pressure bump of width $\Hgas$, \citeauthor{Lin22} estimate the AdSI to dominate the region within a radial distance of $\sim \left(2\alphaM \hgas/\st\right)\Hgas$ from the bump center, which corresponds to $\sim 0.1\Hgas$ for our fiducial parameters, i.e., our simulation domain. Dust trapped close to the center of the pressure bump with $\epsilon\gtrsim 1$ is expected to break into narrow filaments, with the number of filaments increasing with the overall amount of dust trapped. 

On the other hand, the classical SI is expected to dominate unless the radial pressure gradient is sufficiently small. For $\alphaM\sim 0.1$, one expects the classical SI to prevail unless $\widetilde{\eta}\lesssim O(10^{-3})$, approximately an order of magnitude smaller than the canonical value of $\widetilde{\eta}\sim 0.05$. Nevertheless, our results show that a strong accretion flow can boost the maximum $\epsilon$ by a factor of $\sim 3$ than that attainable from the standard SI in non-accreting disks.

In other words, while the AdSI does not dominate under typical midplane disk conditions, an accretion flow lowers the metallicity threshold for strong clumping by the classical SI. We, therefore, expect planetesimal formation to be easier in accreting disks.


\subsection{AdSI ring properties}

\begin{figure}  
    \includegraphics[width=\columnwidth]{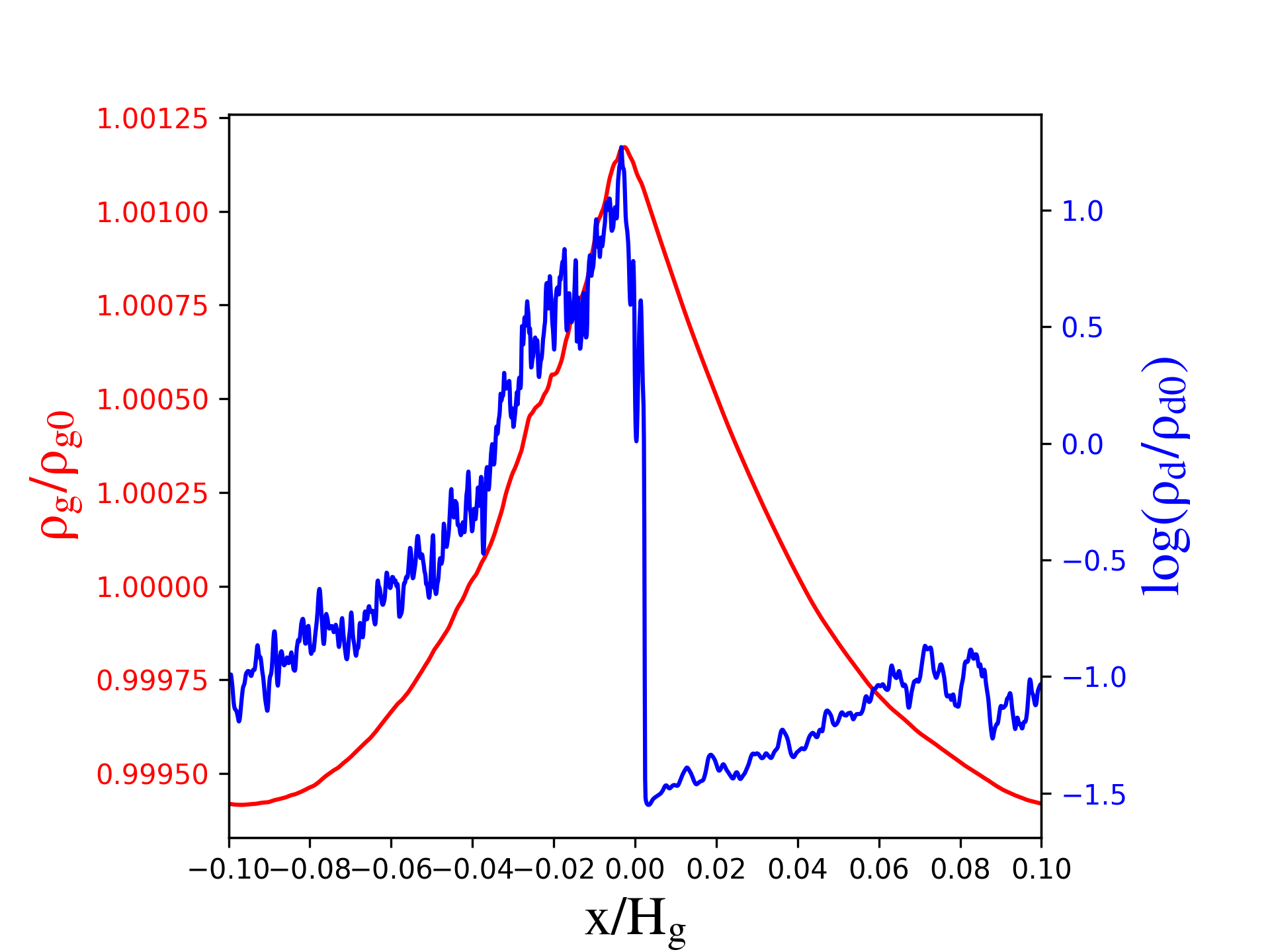}
    \caption{Dust and gas density perturbation of fiducial case at $t=100\ P$. Both dust and gas density strength are normalized by initial density. }
    \label{fig:ringwidth}
\end{figure}

Let us examine the properties of dust rings formed in our simulation. For simplicity, consider the single dust ring that persists at the end of our fiducial simulation ($\widetilde{\eta}=0$, $\epsilon=1$, $\alphaM=0.1$), as shown in the second panel of Fig. \ref{fig:spacetime_eps1}. In Fig. \ref{fig:ringwidth}, we plot the corresponding vertically-averaged radial profiles of the dust and gas densities, which show that dust and gas rings have comparable widths. However, the dust ring is asymmetric about its center. This may be due to the inward accretion flow that works with (against) the inward (outward) drifting dust exterior (interior) to the gas pressure maximum.

To characterize the dust and gas rings, we estimate their widths ($a_\mathrm{d,g}$) as the Full Width at Half-Maximum of their density distributions. We then find $a_\mathrm{d}/a_\mathrm{g} = 0.6$. 

We can compare the above value to the analytical model of dust-trapping by a Gaussian pressure bump developed by \cite{Dullemond18}. Note, however, that the AdSI rings are non-Gaussian, so this comparison is only approximate. Furthermore, their model neglects dust feedback, which is not applicable for our dust rings with $\epsilon\gg 1$. We thus augment it by adding a factor $(1+\epsilon_\mathrm{ch})^2$ in the denominator for the expression for the dust drift velocity \citep[the second term in their Eq. 38, see also][]{Chen20}, where $\epsilon_\mathrm{ch}$ is a characteristic dust-to-gas ratio in the dust ring, treated as a constant for simplicity. Doing so, we find 
\begin{equation} \label{eq:ringwidth}
   \frac{a_{\mathrm{d}}}{a_{\mathrm{g}}}\simeq (1+\epsilon_\mathrm{ch})\sqrt{\frac{\alpha_{\mathrm{g},x}}{\st}},
\end{equation}
where we assumed $\st\ll 1$, and the dust and gas diffusion coefficients are similar. Dust feedback widens dust rings as it slows down radial drift. 

Applying Eq. \ref{eq:ringwidth} with our measured value of $\alpha_{\mathrm{g},x}= 3.13\times10^{-7}$ and taking $\epsilon_\mathrm{ch}\sim 10$, we find $a_\mathrm{d}/a_\mathrm{g}\sim0.02$, still much smaller than that observed in the simulation. This suggests that dust diffusion is much larger than that measured for gas \citep[cf.][]{Youdin07b}. 


\subsection{Implications for ring formation}
A distinguishing feature of the AdSI is that the 100-fold enhancement of dust density --- in filaments --- can be attained even when the initial dust-to-gas ratio is less than unity. By contrast, in this limit, the classical SI only produces mild over-densities without filament formation \citep[][see also Fig. \ref{fig:pureSI_comparison}]{Johansen07}. Thus, the AdSI is a potential mechanism for forming dust rings in dust-poor regions with weak or vanishing pressure gradients. 
Provided that $\epsilon\lesssim 1$ initially, dust rings are not expected to undergo planetesimal formation and persist \citep{Lee24}. On the other hand, AdSI dust rings have narrow widths of $O(10^{-2}\Hgas)$, which cannot be observed directly. An intriguing possibility is AdSI operating within existing dust rings of much larger widths, e.g., those observed by ALMA \citep{Andrews18}. How an underlying AdSI would affect such rings' dynamical evolution remains to be investigated.

\subsection{Caveats and outlook} \label{sec:caveat}

Our models are unstratified, meant to mimic conditions close to the disk midplane. However, in some of our simulations, the resulting vertical dust diffusion coefficients imply a dust layer thickness $H_\mathrm{d}\simeq \sqrt{\alpha_{\mathrm{g},z}/\st} H_\mathrm{g}$ \citep{Dubrulle95} smaller than our vertical domain size of $0.025\Hgas$. For example, our fiducial run with $\alpha_{\mathrm{g},z}\sim 10^{-5}$ and $\st=0.1$ gives $H_\mathrm{d}\simeq0.01\Hgas$. This means the dust layer should exhibit a vertical structure within the domain size. Future models should, therefore, be vertically stratified.

The AdSI filaments in our simulations extend vertically through the domain. It is reasonable to question whether they can form within a finite-height dust layer. However, high wavenumber AdSI (and SI) modes that fit into the dust layer may still operate, provided dissipation is sufficiently weak. Indeed, filaments are readily observed for the SI in stratified simulations \citep{Rucska23,Lim23}. We may thus expect AdSI filaments to persist in stratified disks, but this needs to be addressed explicitly. 

Stratified disks also permit additional drag instabilities related to vertical shear \citep[e.g.][]{Ishitsu09,Lin21} and dust settling \citep[e.g.][]{Krapp20}. These instabilities exhibit weak dust clumping or even dispersal from the midplane. It will be interesting to explore whether a background accretion flow can facilitate clumping in these cases.

Our simulations are also axisymmetric and non-self-gravitating. These approximations must be relaxed to model the breakdown of AdSI—or accretion-induced dust filaments into clumps and their subsequent gravitational collapse into planetesimals. 


Finally, we considered an inviscid disk. However, dust concentration may be suppressed by particle diffusion resulting from external turbulence, for example, driven by hydrodynamic instabilities \citep{Lesur23}. Future simulations can model turbulence as a gas viscosity and dust diffusion \citep[e.g.][]{Chen20}, as a stochastic forcing \citep[e.g.][]{Lim23}, or explicitly account for the underlying instability \citep[e.g.][]{Schafer20}. 


\section{Summary} \label{sec:summary}
This paper presents axisymmetric, unstratified shearing box simulations of dusty PPDs with an underlying laminar gas accretion flow. Previous studies have shown dust interacting with an accretion flow are subject to a new `azimuthal drift' streaming instability (AdSI), distinct from the classical SI \citepalias{Lin22,Hsu22}. 
We extend the first AdSI simulations conducted by \citetalias{Hsu22} across parameter space to examine the effect of initial dust abundance ($\epsilon$), accretion flow strength ($\alphaM$), radial pressure gradient ($\widetilde{\eta}$), and grain size ($\st$).  


Our main findings are as follows:

\begin{enumerate}
\item For $\alphaM=0.1$ and $\st=0.1$, we find the AdSI can enhance dust densities by $100$ for $0.1 \lesssim \epsilon \lesssim 10$. For $\epsilon\gtrsim 1$, the system can thus reach dust densities needed for gravitational collapse, assuming our shearing box is placed in an MMSN-like disk beyond $8\mathrm{AU}$. Interestingly, dust filaments still develop for $\epsilon\lesssim 1$. 




\item AdSI-turbulence yields weak diffusion ($\alpha_{\mathrm{g},x}$) and angular momentum transport ($\alpha_\mathrm{SS}$), with these alpha parameters reaching at most $O(10^{-4})$, even in unrealistically unstable regimes. In our fiducial run with $\alphaM=0.1$, $\st=0.1$, and $\epsilon=1$, we find $\alpha_{\mathrm{g},x}\sim 4\times10^{-7}$ and $\alpha_\mathrm{SS}\sim 10^{-5}$. A stronger angular momentum transport than mass diffusion is typically observed across our parameter study. 

\item An underlying gas accretion flow promotes dust filament formation even when AdSI is formally absent according to linear theory. This is evident in pure SI simulations that should be stable to the AdSI. 
\end{enumerate}

In a follow-up study, we will present simulations of the AdSI in stratified disks, in which the mid-plane dust-to-gas ratio can be self-consistently determined between settling and turbulent diffusion. These will also facilitate direct comparisons with the clumping conditions reported for the classical SI.

\begin{acknowledgments}
We thank the anonymous referee for a clear report and helpful suggestions.
This work is supported by the National Science and Technology Council (grants 112-2112-M-001-064-, 113-2112-M-001-036-, 113-2124-M-002-003-) and an Academia Sinica Career Development Award (AS-CDA-110-M06). Simulations were performed on the Kawas cluster at ASIAA and the Taiwania-3 cluster at the National Center for High-performance Computing (NCHC). We thank NCHC for providing computational and
storage resources. We also thank Chun-Yen Hsu for assisting in calculating correlation times and for providing a reference code.
\end{acknowledgments}

\appendix
\section{Density perturbations} \label{sec:density_perturbation}

\begin{figure*}  
    \includegraphics[width=\textwidth]{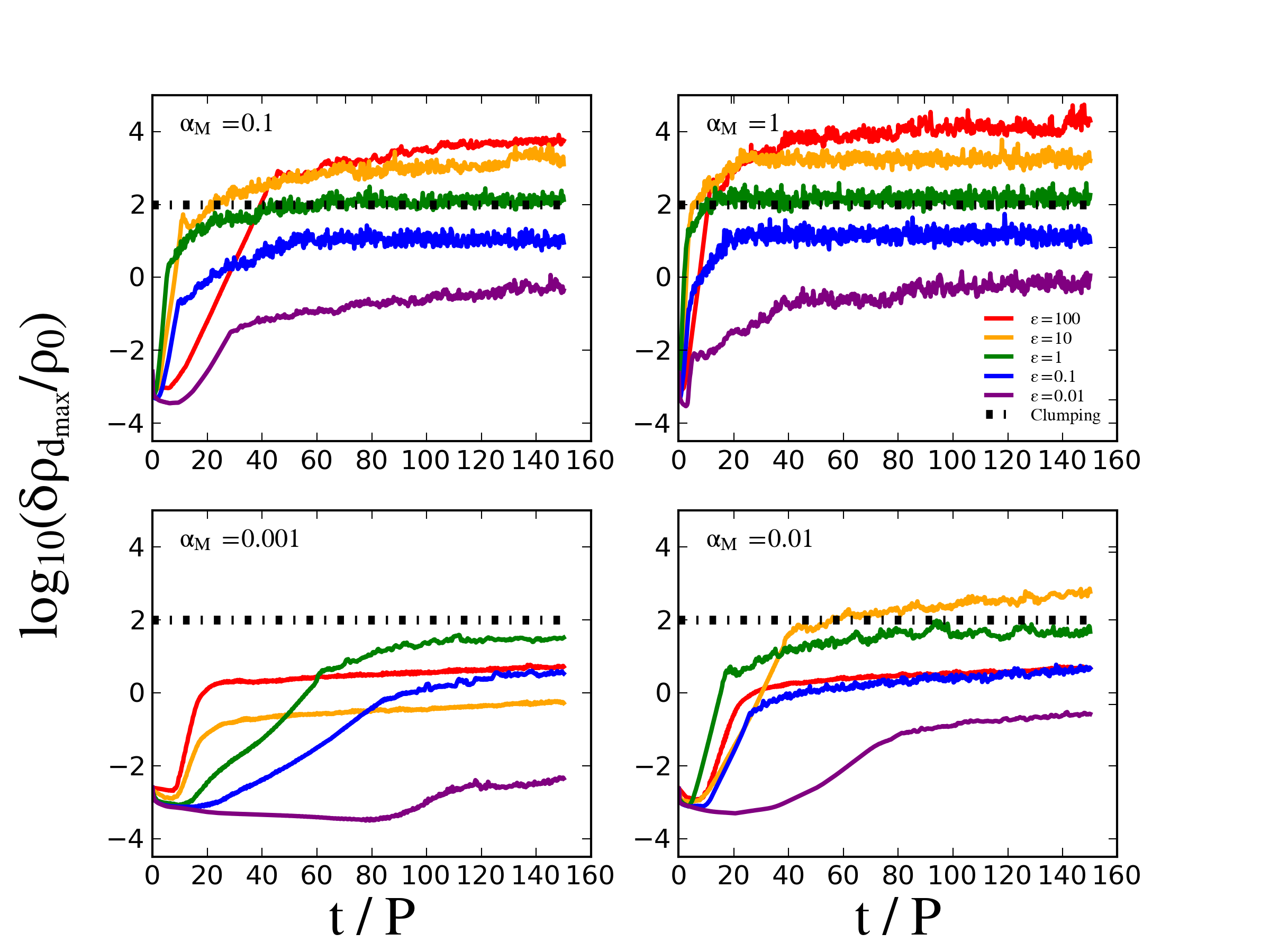}
    \caption{Similar to Fig. \ref{fig:rhodmax_fiducial} but in terms of the maximum dust density perturbation $\delta \rho_\mathrm{d,max}=\operatorname{max}(\rhod-\epsilon \rho_0)$.}
    \label{fig:deltarhod}
\end{figure*}

In Fig. \ref{fig:deltarhod}, we re-plot the dust density evolution shown in Fig. \ref{fig:rhodmax_fiducial} and Fig. \ref{fig:rhodmax_varyam} in terms of the maximum dust density perturbation, i.e. $\delta \rho_\mathrm{d,max}=\operatorname{max}(\rhod-\epsilon\rho_0)$. This more clearly illustrates the instability growth at early times. For $\alphaM\geq 10^{-2}$, the maximal growth occurs for $\epsilon=1$ (green curve), consistent with linear theory (\S\ref{linear}). This trend is not reproduced for $\alphaM=10^{-3}$, likely due to the difficulty in capturing slowly-growing modes in these stable cases. 


\section{Box size selection}
\label{sec:boxsize}

\begin{figure*}  
    \includegraphics[width=\textwidth]{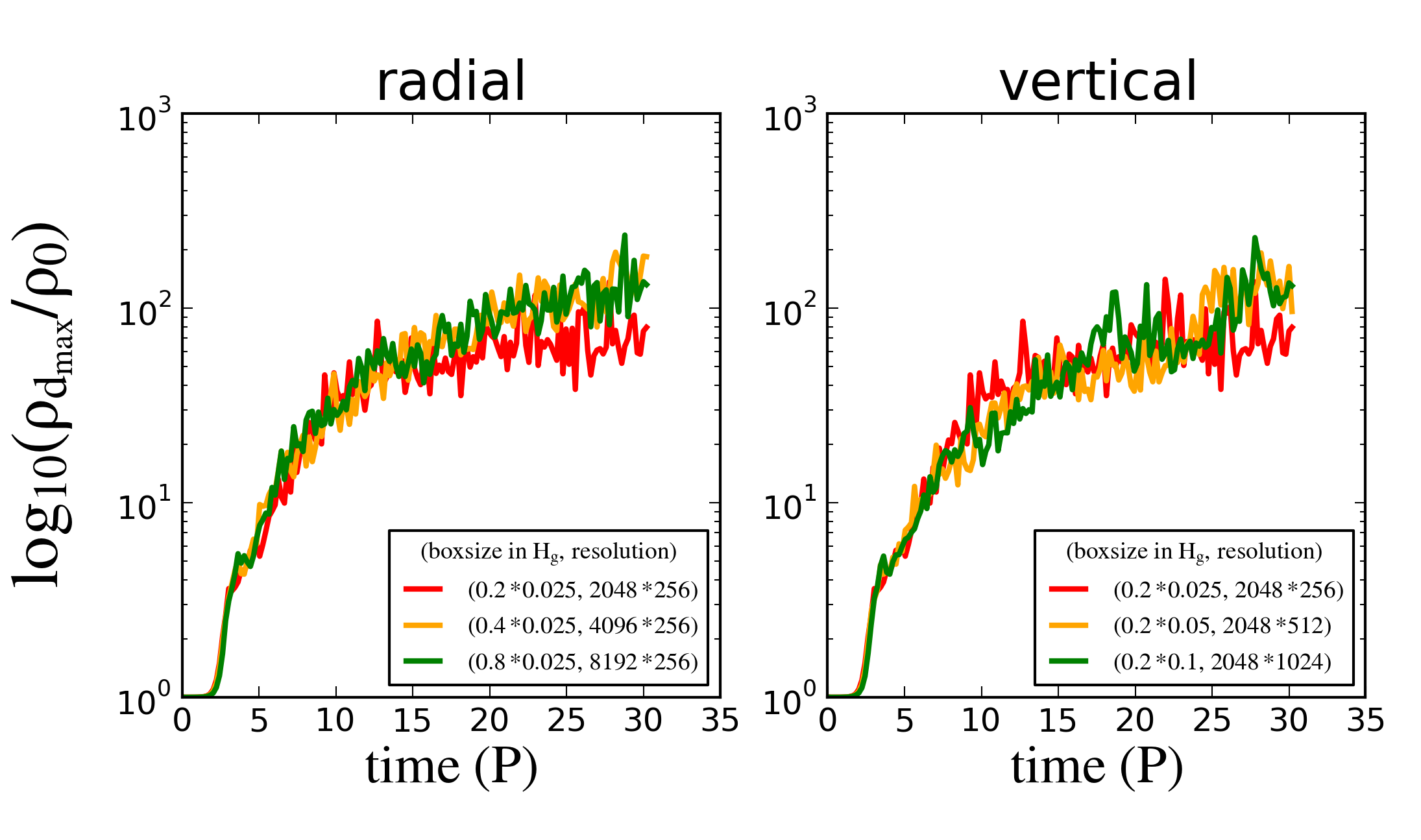}
    \caption{Evolution of maximum dust density with different box sizes. We compare the simulation box in this study (red line) with the box twice and four times the length along each side. The left and right panels show the changing box size in a radial and vertical direction. The number of cells is chosen to keep the resolution per $\Hgas$ constant.} 
    \label{fig:boxsize}
\end{figure*}

Our physical setup is similar to \citetalias{Hsu22} but with half the vertical domain size. In Fig. \ref{fig:boxsize}, we perform additional, short runs for the fiducial parameters ($\st=0.1$, $\alphaM=0.1$, $\epsilon=1$, $\widetilde{\eta}=0$) with larger radial and vertical domain sizes in the left and right panels, respectively. Our fiducial run is shown in red, while double and quadruple domains are shown in orange and green, respectively. We find larger dust densities with increasing domain size, by about a factor of two. Convergence may be reached with twice the domain size, but the associated computational cost becomes unfeasible for a parameter survey with longer simulation times. We thus chose the smaller domain as a compromise. 



\bibliography{main}{}
\bibliographystyle{aasjournal}



\end{document}